\documentclass{sig-alt-full}
\usepackage{subfigure}
\usepackage{times}
\begin{document}

\title{Avatar Mobility in Networked Virtual Environments:\\
Measurements, Analysis, and Implications}
\numberofauthors{2}
\author{
\alignauthor Huiguang Liang, Ian Tay, Ming Feng Neo\\
\vspace{1mm}
    \affaddr{Dept. of Electrical and Computer Engineering}\\
    \affaddr{National University of Singapore}\\
    \and
\alignauthor Wei Tsang Ooi\\
\vspace{1mm}
    \affaddr{Dept. of Computer Science}\\
    \affaddr{National University of Singapore}\\
\alignauthor Mehul Motani\\
\vspace{1mm}
    \affaddr{Dept. of Electrical and Computer Engineering}\\
    \affaddr{National University of Singapore}
}
\maketitle
\begin{abstract}
We collected mobility traces of 84,208 avatars spanning 22 regions
over two months in Second Life, a popular networked virtual
environment.  We analyzed the traces to characterize the dynamics
of the avatars mobility and behavior, both temporally and spatially.  We discuss
the implications of the our findings to the design of
peer-to-peer networked virtual environments, interest management, mobility modeling of
avatars, server load balancing and zone partitioning, client-side
caching, and prefetching.  
\end{abstract}


\section{Introduction} \label{sec:intro}

The possibility of multiple users communicating and interacting with
each other in a networked virtual environment (NVE) over the Internet 
has excited many researchers in the past 20 years.  
Building an NVE that scales to many users, while
maintaining interactivity requirements, however, remains a difficult 
challenge.  Much research effort has gone into reducing communication
overhead, maintaining state consistency, and managing server resources.  

These research efforts were previously handicapped by a lack of
deployed, open, large-scale NVE systems, on which the researchers
could evaluate 
the effectiveness of their proposed solutions.  In particular, the
effectiveness of many of these solutions depends heavily on the avatar
behavior, movements, and interactions within the NVE.  Without such
data, previous research mainly based their evaluations either on simulations 
with a simple model of avatar behavior (such as random way-point 
mobility model), or on collected traces from small-scale games and NVEs.

In recent years, however, the increase in bandwidth
to home users and availability of powerful graphics capability in
commodity PCs have lead to development of several NVEs
targeted at the Internet mass.  Among the notable NVEs
are Second Life, There, Active Worlds, and HiPiHi.
Second Life is the most popular NVE
available, with an average of 38,000 simultaneous users spending 
28 million man-hours in January of 2008
alone\footnote{http://secondlife.com/whatis/economy\_stats.php}.  
Furthermore, the Second Life client is open source, providing
opportunities for reverse
engineering the protocols on which Second Life runs.
We believe that the availability of such large-scale, open NVEs 
provides exciting opportunities for researchers to 
evaluate their solutions using  a large amount of real traces under 
realistic scenarios.  This belief drives our work in this paper.

This paper presents our effort in collecting and analyzing avatar
traces from Second Life.
By using a custom Second Life client, we collected the identity, action, positions, and 
viewing directions of 84,208 avatars over two months,
spanning 22 regions on Second Life, giving a total of 62 million
records.

We analyze our traces to study the temporal and spatial dynamics of
avatars.
For temporal dynamics, we ask the following questions:
(i) How does avatar population vary over time?  (ii) How long does
an avatar stay in a region? (iii) How often does an avatar return to
the same region?  (iv) If an avatar does return, how much time has
passed before the avatar returns? and (v) How do avatar arrival and
departure rates vary over time?

The traces also provide a rich amount of information on the spatial
distribution of avatars in a region and their movement patterns.  
We divide the region into cells, and ask: Given a cell,
(i) how 
many times does an avatar visit the cell?  (ii) How long does 
an avatar stay in the cell? (iii) How fast does an avatar move 
in the cell?  

We also analyze the contact patterns
among the avatars.  In particular, we are interested in characterizing
(i) the number of avatars within an avatar's area-of-interest, 
(ii) the duration two avatars stay within each other's
area-of-interest, and (iii) how dynamic is the set of avatars within
an avatar's area-of-interest.

Our traces and analysis are useful in many ways.  
First, the traces, which we plan to share with the research 
community, capture actual movements and activities of 
large number of avatars.
It can be used in trace-based simulations of NVEs, allowing
new and existing designs and algorithms to be evaluated under
realistic conditions.  

Second, the traces can be used to derive and verify new mobility 
models for NVEs.  Most research assumes a mobility model
based on random walk \cite{Makbily99,Vik06}, random waypoint or 
their variations \cite{Chertov06,Chen06,Backhaus07,Bharambe04}.  
Research, however, has shown that the evaluation results based on
these simple models are significantly different from those based
on actual traces.  A new avatar mobility model for NVE is therefore needed.
Our traces and findings serve as a crucial first step towards that
goal.
The traces can also help in deriving and verifying appropriate models for
the spatial distribution of avatars in the virtual world, where past
research has assumed that avatars are uniformly distributed \cite{Lu04}
and distributed in clusters \cite{Lui02}.

Finally, our analysis provides insights into how avatars behave
and move in an NVE such as Second Life.  This knowledge
can lead to design of new and more effective algorithms for NVEs.
For instance, in designing a peer-to-peer NVE, it is important to
understand the expected churn rate, identify peers that stay
in the system for a long time, and understand if (and how) the
avatars move and congregate.  In designing load balancing and zone
partitioning schemes for NVE servers, knowing
the expected spatial distribution of avatars and their tendency to
moves across zones is helpful.

It is not our intention in this paper to evaluate previous work
using our traces, nor do we intend to propose new mobility and
avatar behavioral models.  These are important research directions that we
believe need to be explored, but do not fit into the scope
of this paper.  Instead, in this paper, we discuss our findings from
our analysis of the measured traces,
and how the findings will affect various aspects of NVE
design in general.  

The rest of this paper is structured as follows.  We present previous
work related to ours in Section \ref{sec:related}.  Section
\ref{sec:secondlife} briefly introduces Second Life.  We
explain how we collect and verify our data in Section
\ref{sec:collection}.  Section \ref{sec:analysis}
presents our analysis of the traces.  
We discuss the implications of our traces in Section
\ref{sec:implications}.  
Finally, we conclude in Section \ref{sec:conclude}.

\section{Related Work} \label{sec:related}

We now describe previous efforts in collecting avatar traces
from networked virtual environments and games.
Rieche et al. \cite{Rieche07} collected a 5-hour trace of 400 players
from an online game called FreeWar.  
Boulanger et al. \cite{Boulanger06} collected a trace of 28 
players from a game they developed called Orbius.  The focus of
their work is not on the trace, but rather, the trace is a 
way to evaluate their proposed algorithms.  Rieche et al. use
their trace to evaluate a load balancing scheme, while Boulanger et
al. use their trace to evaluate different interest management
algorithms.  Beside traces collected from games, both works 
use randomly generated movements in their evaluation, and both 
observe significant
differences in their results evaluated using the traces and using
generated movements.  Their results highlight the importance of
having real mobility traces for researchers to evaluate their work.

Tan et al. \cite{Tan05} and Bharambe et al. \cite{Bharambe06} collected traces
from Quake III, a popular, multi-player, first person shooting (FPS) game
and developed mobility models to describe the movement of the 
players.  Pittman et al. \cite{Pittman07} collected a large trace,
comparable in scale to ours, of players movement from World of 
Warcraft (WoW), a massively multi-player online role playing game (MMORPG), and
analyzed the
dynamics of the populations, players arrival/departure rate, session
length, player distribution, and player movements.  FPS games and MMORPGs
have different characteristics than NVEs.  Players in fast-action, FPS games 
tend to move around constantly.  In MMORPGs, 
players usually engage in quests to gain level and new abilities.
Players tend to gather in a location for an event (e.g.
new monsters to fight) and disperse afterwards.  Players also tend
to move in groups.  We observed a different pattern for NVEs.

Most recently, La and Pietro have independently conducted a similar 
study on mobility in Second Life \cite{La08}.  Their study, however,
focuses on metrics relevant to mobile communications, such as 
graph theoretic properties of line-of-sight networks formed by the 
avatars, travel length and time of avatars, and contact opportunities
among avatars.  Their goal is to use the mobility traces of avatars
to model human mobility for applications related to wireless
and delay-tolerant networks.  On the other hand, we focus on
metrics that are of interest to systems design of NVEs.

\section{Second Life} \label{sec:secondlife}

Before we describe how we collected our traces, we briefly 
introduce Second Life in this section.  Second Life is an
NVE launched by Linden Lab in 2003.  
It is a so-called \textit{metaverse}, where users participate
in creating the virtual world by constructing buildings 
and authoring objects.  Furthermore, users control avatars 
that can interact with each other, socialize, and trade user-created
objects. 

Unlike popular massively
multiplayer online games (MMOG) such as World of Warcraft, the 
virtual world in Second Life is highly dynamic -- users can create 
objects, place them into the virtual world, and write scripts
to program the behavior of the objects.  In comparison, the world
or game maps in MMOG are mainly static and are built by game
publishers.  As such, it is possible to distribute the data describing 
the game world on DVDs.  In Second 
Life, however, only a viewer program is distributed to the users. 
Data pertaining to the virtual world, such as terrain, objects, 
behavioral scripts, and textures, are downloaded on demand as the
user explores and interacts with the virtual world.

The virtual world in Second Life
is made up of \textit{regions}.  Each 
region is a 256m $\times$ 256m piece of land,
managed by a Second Life server process that
maintains the states of all avatars
and objects within the region.
The virtual world in Second Life is not seamless -- the user 
cannot walk seamlessly between regions, but rather
has to \textit{teleport} from region to region.  As the user
teleports, the state of the avatar is transferred to the 
destination's simulator (usually takes order of seconds).  
Each region has a teleportation point called the \textit{landing
point}, where all arriving avatars will appear.

An activity unique to Second Life is \textit{camping}, where
avatars can earn free virtual money by engaging in certain
activities (e.g., get paid by the hour to sit 
on a chair)  Region owners typically use camping to boost the
popularity of the regions.

Within a region, a user can walk, run, or fly.  The user
can also teleport from one place to another within a region, if
intra-region teleportation points exist.
In this paper, we are interested in 
collecting the mobility traces of the avatars as they move around 
within a region.  We describe how we collect the traces next. 

\section{Data Collection} \label{sec:collection}

To collect the traces of avatars, we developed a client for
Second Life based
on an open source library called
libsecondlife\footnote{www.libsecondlife.org}.  
Our client 
visits selected regions in Second Life using a bot, and by
parsing update packets from the servers, we can obtain information
about other avatars in the regions.  We 
log information about all detected avatars at ten second intervals.

Second Life uses an architecture that stores the states of all avatars 
in a centralized server.  The server pushes information about other
avatars and objects in the region to the clients, depending on the 
client's avatar position.  How the server decides which information to
push is unfortunately proprietary and unknown to us.  It was therefore 
not clear to us whether our bot is able to track the positions of every
avatar on the region.  For instance, if the server uses interest
management techniques, then only positions of other avatars within the
area-of-interest (AoI) of our bot will be updated, while those outside are
culled.

We inserted several avatars 
into Second Life to determine how interest management
is done.  We observe that culling indeed does occur for avatars
outside of the AoI, but, once the bot starts detecting 
an avatar, updates about that avatar will no longer be culled, even
if the avatar moves outside the bot's AoI.  To
track the movement of as many avatars as possible,
we therefore place our bot at the landing point.
As such, all incoming avatars will be immediately
detected by the bot and subsequently our bot will receive update
of these avatars regardless of where they are in the region.
The existing avatars
that are already in the region when we started our bot, however, may not be
tracked.   Fortunately, our measurements show that 95\% of the avatars
stay less than an hour in the region (see Section
\ref{sec:sessionbehavior}).  Thus, after the first couple of hours,
our bot should be able to track nearly all of the avatars in the
region.

\subsection{Difficulties Encountered}
We now briefly explain several issues we ran into 
during our data collection process,
which lead to imperfect traces.  

One source of imperfection in our traces is related to avatars that
are sitting on objects.  Second Life reports the locations of
these avatars relative to those objects.  To recover the position
of the avatars, we therefore need to know the position of the objects
they are sitting on.  We observe that packets containing 
information on objects are sometimes culled if they fall outside 
the bot's AoI.  In this case, we are unable to compute the position
of the avatars sitting on these objects.  Such records are removed from 
the trace, but only when we compute metrics that require the positions 
of the avatar.  
About 8\% of our records have unknown avatar positions.

Occasionally, our bots are
kicked out by the region owner, as the owner may not take
kindly to bots.  Each region allows a
limited number of avatars inside at any one time, due
to server's resource constraints. An inactive bot occupies a valuable
slot that could have been filled by an actively contributing avatar.
On a few occasions, region owners went as far as to ban our bots from
returning, creating temporal breaks within the data set.

 Temporal breaks also happen when our client
 crashes due to insufficient memory.  The libsecondlife library maintains
 a staggering amount of states about each object and avatar in a region
 (at least 1 GB in densely populated regions).
 We have a script to automatically restart the client in such situations, but
 the bot may not be able to log back to the same region, because the
 region becomes full during the crash, creating temporal breaks in our traces.

\subsection{Verification}

Positional prediction techniques, such as dead reckoning, are commonly
used in NVE to reduce update frequency between server and clients at
the cost of reduced consistency in avatar position.  We are concerned
with how consistent the reported positions of the avatars are and if
any predictions are done at the client. 

An experiment was conducted to verify the consistency of the
reported position.  We placed seven bots on a region called Freebies,
with four static bots at each corner of the region, one static bot
at the center of the region, and two bots that walk around
the region following a random walk model.  We log the positions of
other avatars in the region as seen by each bot.  For each avatar
at each time instance, seven records of the positions are obtained.
We compute the standard deviations of x and y positions
of the avatars seen by these bots.  We found that only 1.14\% of
the reported x positions and 1.39\% of the reported y positions 
have a standard deviations of more than 10 meters, while  9.00\% and 
10.15\% of the records have standard deviations of more than
1 meters for x and y positions respectively.  There are instances
where the two bots detected the same avatar with reported positions
more than 100 meters apart -- this happens when the avatar teleports
with the region.  Delay in receiving updates from the 
server causes these discrepancies.

Another concern we have is whether our bot consistently observe the
same set of avatars.  To verify this, we place two bots 
close together at the teleportation point on Freebies, and each bots
recorded the number of avatars it currently knows of every 
10 seconds for 30 minutes.
We found that the numbers reported by the two bots are very close. 
Each bot detected an average of 71.22 avatars, and the mean difference 
between the reported number of avatars at each scan is 1.16 (1.63\%).  
This difference is caused by one to two avatars
(who are already in the region when we started our experiments)
moving into the AoI of one bot but not the other.

The verification experiments above indicate that our collected data 
contain some errors, caused by state synchronization delay and
interest management techniques used by the server.  With 
no access to the server states, we can
only collect the traces at a client, where these errors
are unavoidable.  Fortunately, we found that the errors are
reasonably small and we believe they will not affect the general
conclusions we obtain from the analysis of the traces.

\subsection{Limitations}
\label{sec:limitations}

Our traces and analysis have two limitations.
First, our bot cannot tracks movement of avatars 
between regions.  Second Life
does not provide information about the destination of an avatar
when it leaves a region.  In fact, we cannot differentiate
between an avatar logging out and teleporting to another region.

Second, we ignore the z-coordinates of avatars in our analysis.
Each avatar position gives the coordinate of the avatar in a
3D space.  We observe, however, that most of the time the avatars
stay on the ground.  We therefore 
focus only on the x- and y-coordinates.  As a result,
an avatar that hovers in the air is considered to be at the same
position as another avatar standing on the ground if they have
the same x- and y-coordinates.

\subsection{Regions}

\begin{table}[hbt]
\center
\begin{tabular}{r|r|r}
Name & Number of Avatars & Date \\
\hline
Isis & 2,735 & 28 Mar 2008, Fri \\
Ross &  560 & 11 Mar 2008, Tue \\
Freebies & 3,153 & 11 Mar 2008, Tue\\
The Pharm & 1,537 & 5 Mar 2008, Wed\\
Isis (Long) & 8,795 & 28--31 Mar 2008, Thu to Sun\\
\end{tabular}
\caption{Summary of Traces Analyzed\label{tab:regions}}
\end{table}

While we collected data from 22 regions over several weeks,
we focus our analysis on a one-day trace from four 
regions in this paper, namely \textit{Isis}, \textit{Ross},
\textit{Freebies}, and \textit{The Pharm}.  
A summary of these traces are shown in Table \ref{tab:regions}.

We focus on three popular regions, Isis, Freebies, and 
The Pharm.
Isis has a mature adult theme.  Residents can participate 
in paraphiliac activities, buy adult-novelty items, and camp.  
Freebies gives away free objects, clothes, accessories, and 
other inventory items to any resident. It also features a 
very small camping area.  The Pharm is a region focus only
on camping.  These three regions are consistently among the
most popular regions in Second Life.

Ross, a region with medium popularity is chosen 
to contrast the results from the popular regions.
Ross is provided by Linden
Lab to distribute information and serves as meeting
place for avatars.

Besides popularity and variation in themes, we choice of
these four regions is also due to the
completeness of their traces.  As mentioned in Section
\ref{sec:limitations}, we encountered temporal breaks in our traces.
The set of traces from these regions are the most complete,
with only six, small breaks in between.  The average break
is 8 minutes, with the longest break being 16 minutes and 30 
seconds.  Out of four days, only 0.8\% of the traces are lost 
due to the breaks.

Besides the above traces, we also included our
analysis for a 4-day trace for Isis to study if there is
any changes in our observation over multiple days, 
covering both weekdays and weekends.

We did not choose any unpopular regions to study, as those 
usually yield few avatars and give no interesting results.

\section{Characterizing Avatars}
\label{sec:analysis}
\subsection{Session Behavior} 
\label{sec:sessionbehavior}

\begin{figure*}[tb]
\centering
\subfigure[]{
	\label{islands_num_of_avatars_over_t}
	\includegraphics[width=.31\textwidth]{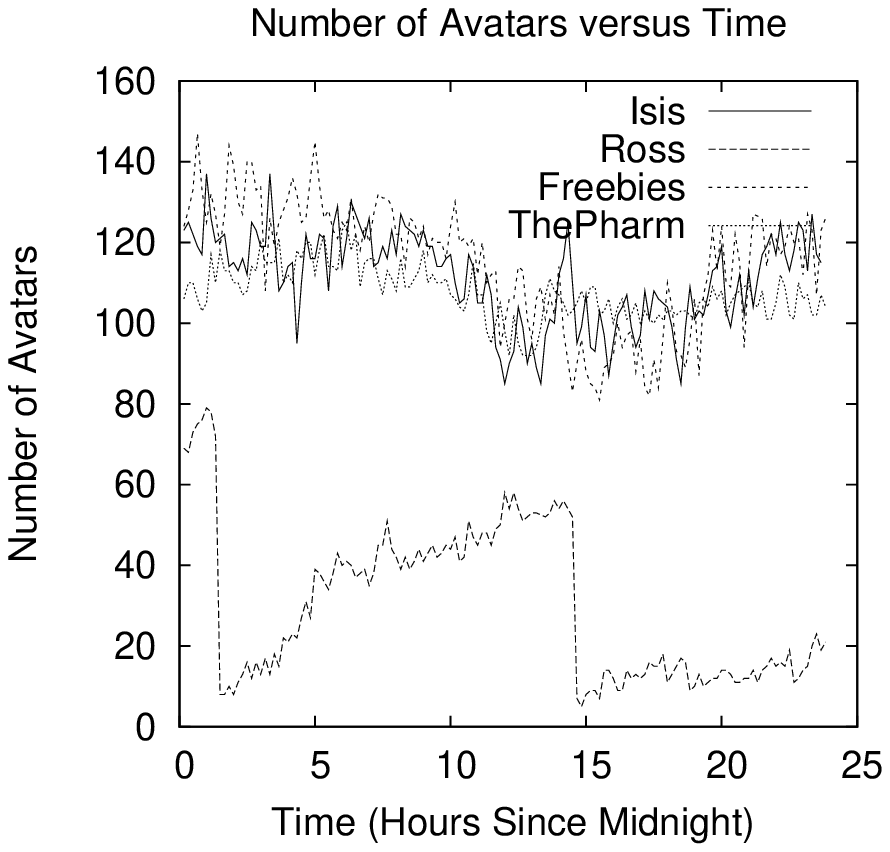}}
\subfigure[]{
	\label{islands_num_of_arrivals}
	\includegraphics[width=.31\textwidth]{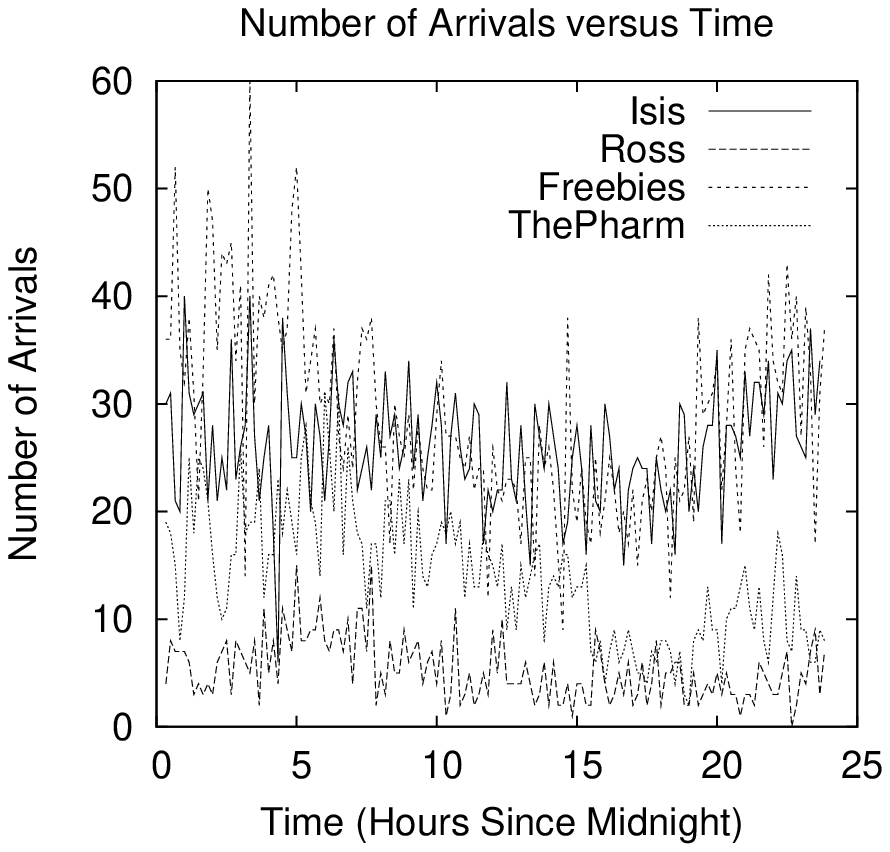}}
\subfigure[]{
	\label{islands_num_of_departures}
	\includegraphics[width=.31\textwidth]{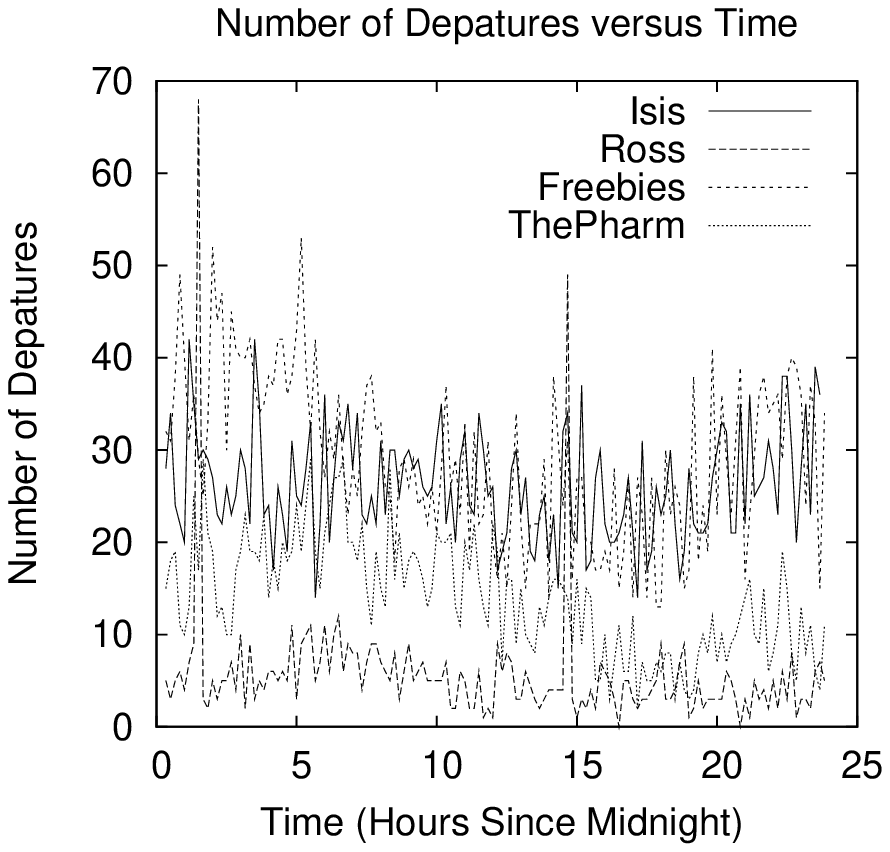}}
\subfigure[]{
	\label{islands_num_of_returns_cdf}
	\includegraphics[width=.31\textwidth]{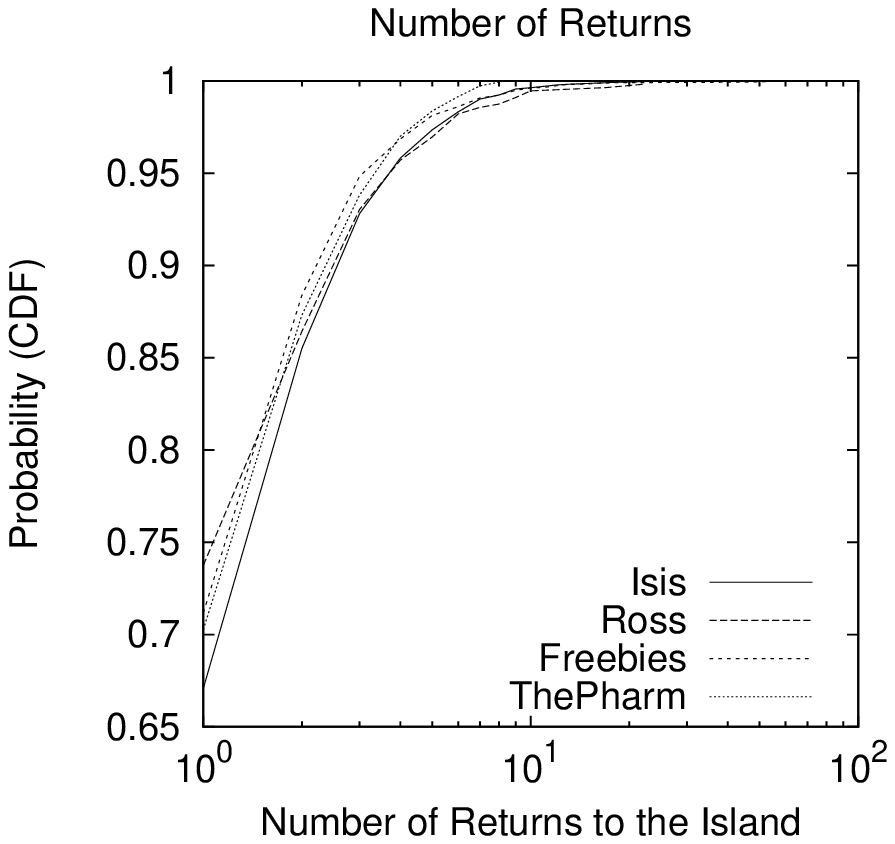}}
\subfigure[]{
	\label{islands_inter_return_time_cdf}
	\includegraphics[width=.31\textwidth]{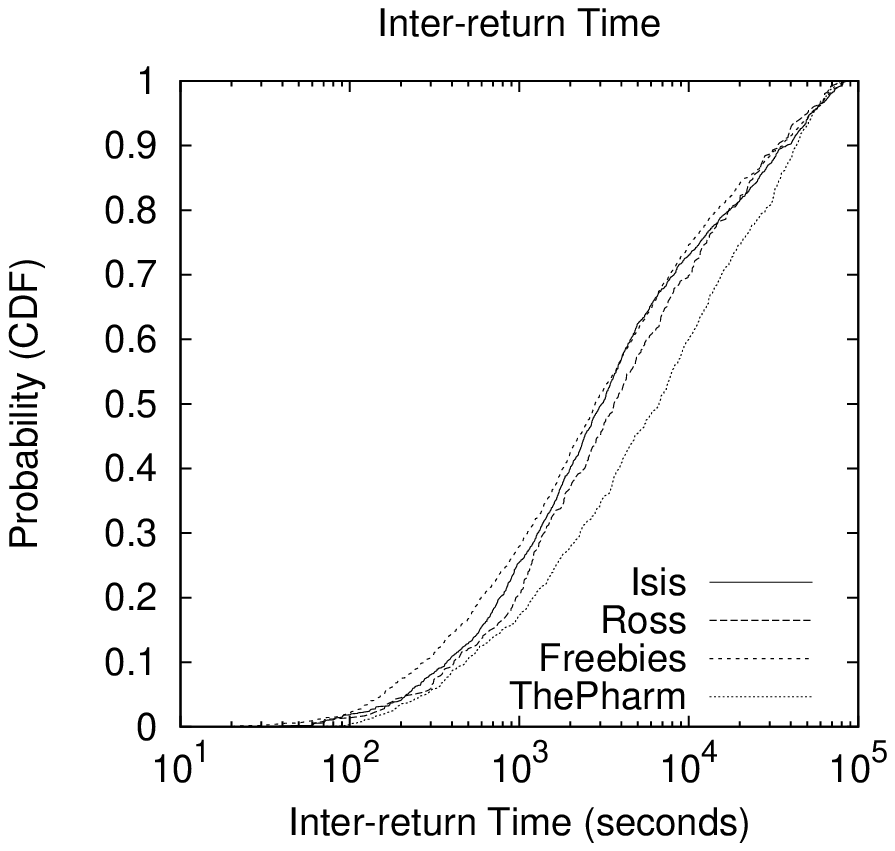}}
\subfigure[]{
	\label{islands_stay_time_cdf}
	\includegraphics[width=.31\textwidth]{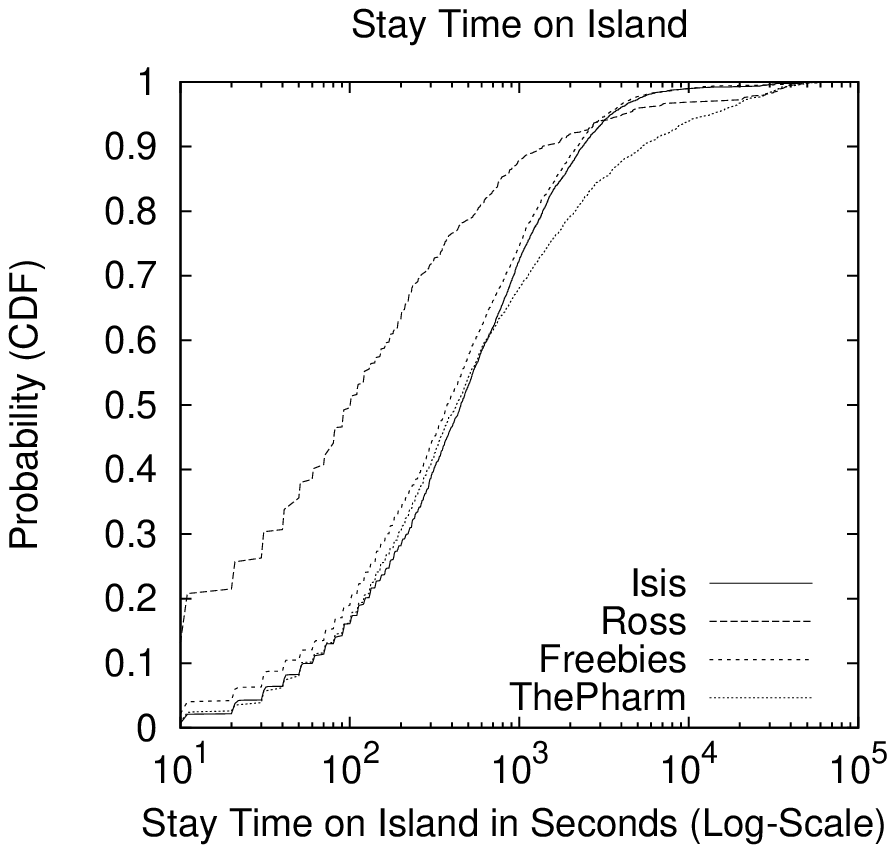}}
\caption{Session Behavior}
\end{figure*}

We now report on the session behavior of avatars based on
the traces in Table \ref{tab:regions}.

\textbf{Population over Time:} Figure
\ref{islands_num_of_avatars_over_t} shows how region population
changes over time.   The three popular regions, Isis, Freebies, and 
The Pharm exhibits similar patterns in population over 
time where there is a slight drop in population from 12 noon to 6 pm.
The diurnal pattern is not as obvious as observed in other NVEs
\cite{Pittman07}.  The population from Ross exhibits a different pattern, where
population increases steadily but suddenly drops.  These drops could be
due to server reset, causing all avatars (include our bot)
to be logged out.  
We observe similar patterns when studying our traces from other days.
Our 4-day Isis trace, for instance, exhibits similar drop,
surprisingly, even on weekends.

We have two possible explanations for the lack of obvious diurnal
pattern. First,
Second Life limits the number of avatars visiting the regions.
For popular regions, there is a constant demand for users to enter
the region -- so, when a user leaves and the region has a slot open,
another users is likely to enter the region.
Second, Second Life users span many parts of the world.  According
to the economy statistics published by Linden Lab, about 40\% of 
the users come from North America. Another 40\% come from Europe. 
The rest are from Asia, South America, and Oceania.  Users log onto
Second Life at all times of the day.  The slight drop during 12 noon
to 6 pm local time corresponds to daytime in North America.  The
diurnal pattern caused by users in Europe is not as obvious as the
continent spans many time zones.

\textbf{Arrivals and Departures:} Figures \ref{islands_num_of_arrivals} and
\ref{islands_num_of_departures} show the number of arrivals and
departures over time.  These figures show that the population on 
the regions is highly dynamic, as expected from the maximum number 
of avatars in a region (Figure \ref{islands_num_of_avatars_over_t})
and number of unique avatars observed (Table \ref{tab:regions}).  We 
can see a high churn rate for Freebies, as avatars tend to drop into
the region, picks up free items, and leaves.  The churn rate is
especially high from 12 midnight to 6am (up to 60 churns per
hour).  The churn rates for The Pharm and Ross are lower, 
due to the camping activities and low popularity of the region, 
respectively.

\textbf{Returning to the Same Island:} We are interested in how many times an avatar
revisits a region.  Surprisingly, even within a day, we observed
multiple visits by the same avatar.  The maximum number of revisits
observed is 55 (for Freebies).  We speculate that these avatars might be
hoping from region to region.  About
25 - 35\% of the avatars revisited the same region within a day.
The CDF for this metric is shown in Figure
\ref{islands_num_of_returns_cdf}.   Figure
\ref{islands_inter_return_time_cdf} shows the CDF for the time 
that passed between an avatar leaving the region and returning
to the region. We call this metric \textit{inter-return time}.
The median inter-return time for the three regions on the day observed
ranges from 45 minutes to an hour.  90\% of the inter-return time
observed is less than 10 hours.  

\textbf{Stay Time:} Figure \ref{islands_stay_time_cdf} shows the cumulative distribution
of how long an avatar stays in a region.  We call this 
the \textit{stay time} of an avatar.  We note
that this does not correspond to the time an avatar stays in Second
Life, since the avatar could have just teleported to another region 
rather than leaving Second Life.  We compute the stay time by logging
the time between the arrival and departure of an
avatar, excluding all avatars that are already in the region at the
beginning of our trace.  The distribution of stay time is 
highly skewed, close to a power law distribution.  An obvious
observation is that the stay time at Ross is lower (median of 92
seconds) than that of Isis and Freebies (median of 448 seconds and 373
seconds).  The distribution of stay time at The Pharm is skewed 
towards higher values (despite a median of only 427 seconds) than 
Isis and Freebies, since avatars have incentives to stay on the
island.  The periodic reset on Ross could explain the shorter stay time.  

\subsection{Mobility} \label{sec:userbehavior}

\begin{figure*}[p]
\centering
\subfigure[Isis]{
	\label{isis_num_of_visits_map}
	\includegraphics[width=.22\textwidth]{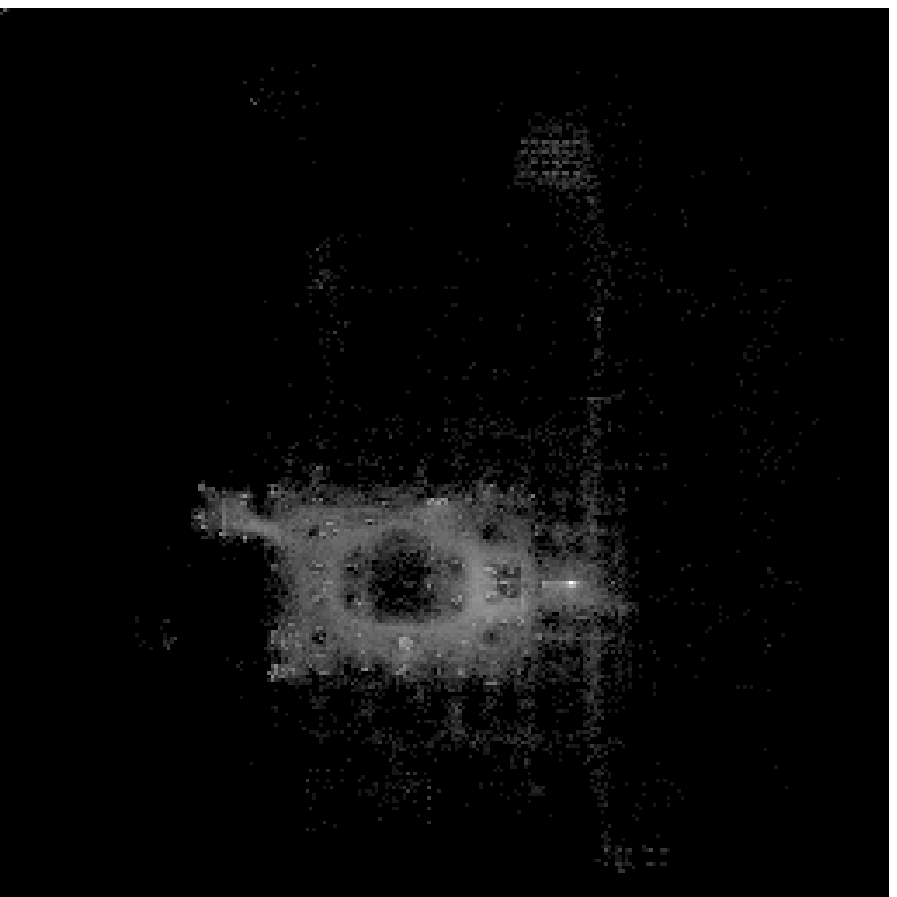}}
\subfigure[Ross]{
	\label{ross_num_of_visits_map}
	\includegraphics[width=.22\textwidth]{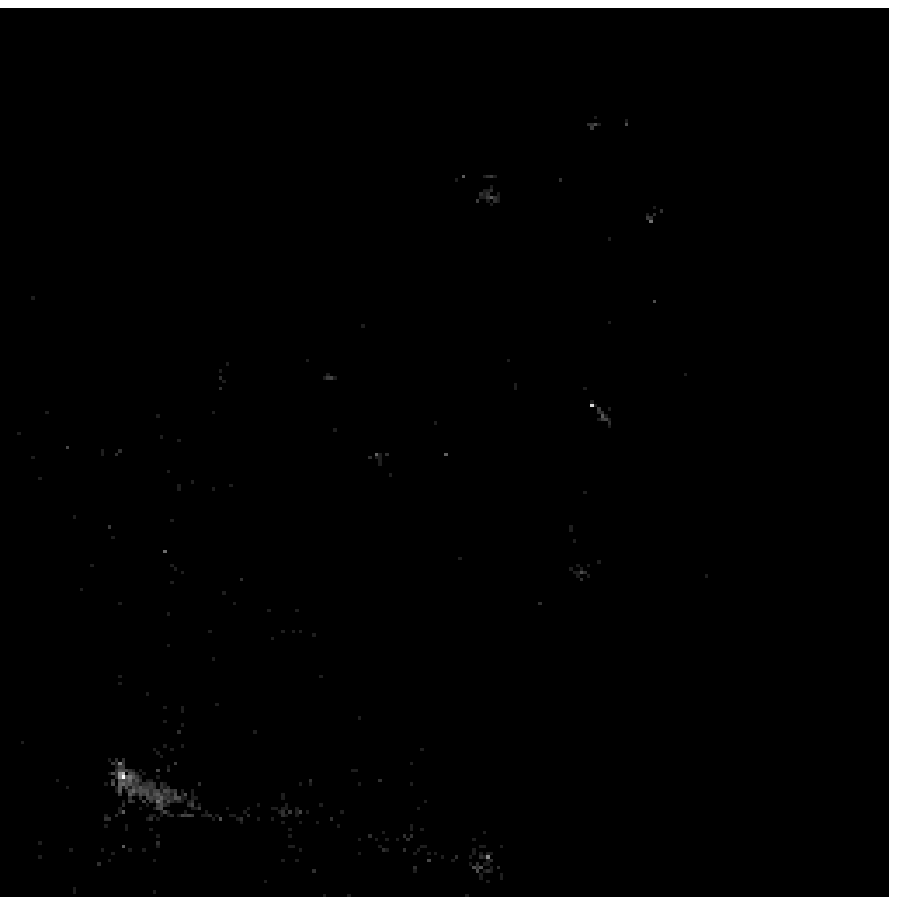}}
\subfigure[Freebies]{
	\label{freebies_num_of_visits_map}
	\includegraphics[width=.22\textwidth]{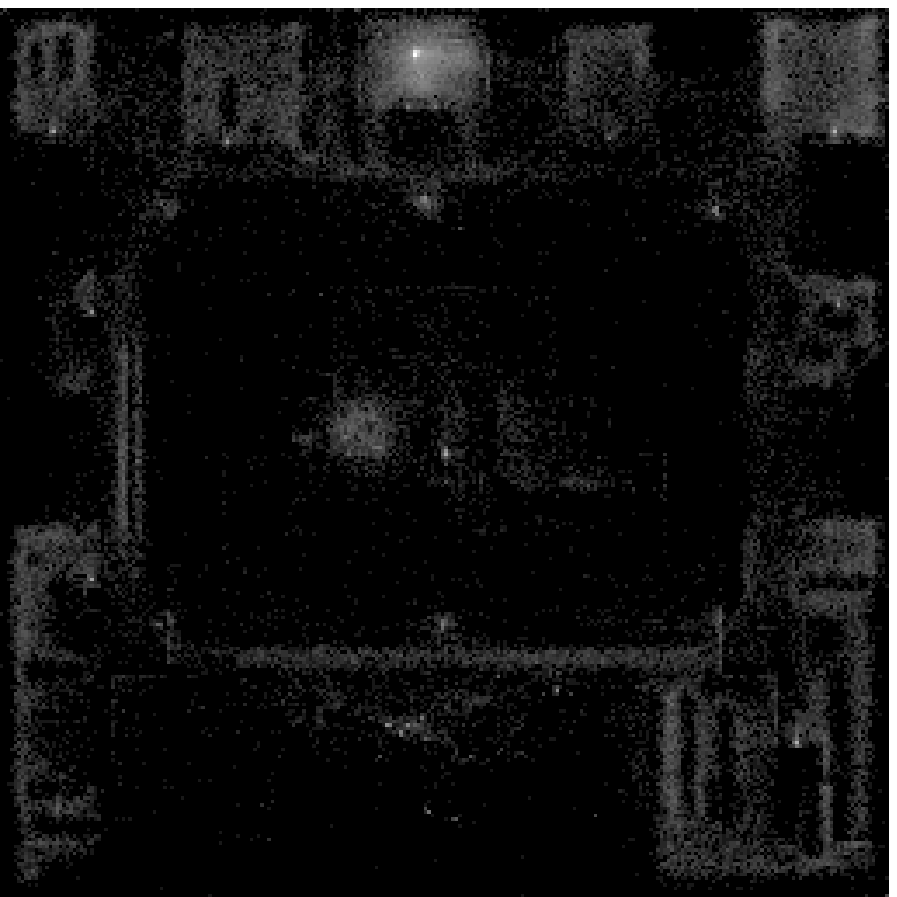}}
\subfigure[The Pharm]{
%\subfigure[The Pharm: Number of Visits]{
	\label{pharm_num_of_visits_map}
	\includegraphics[width=.22\textwidth]{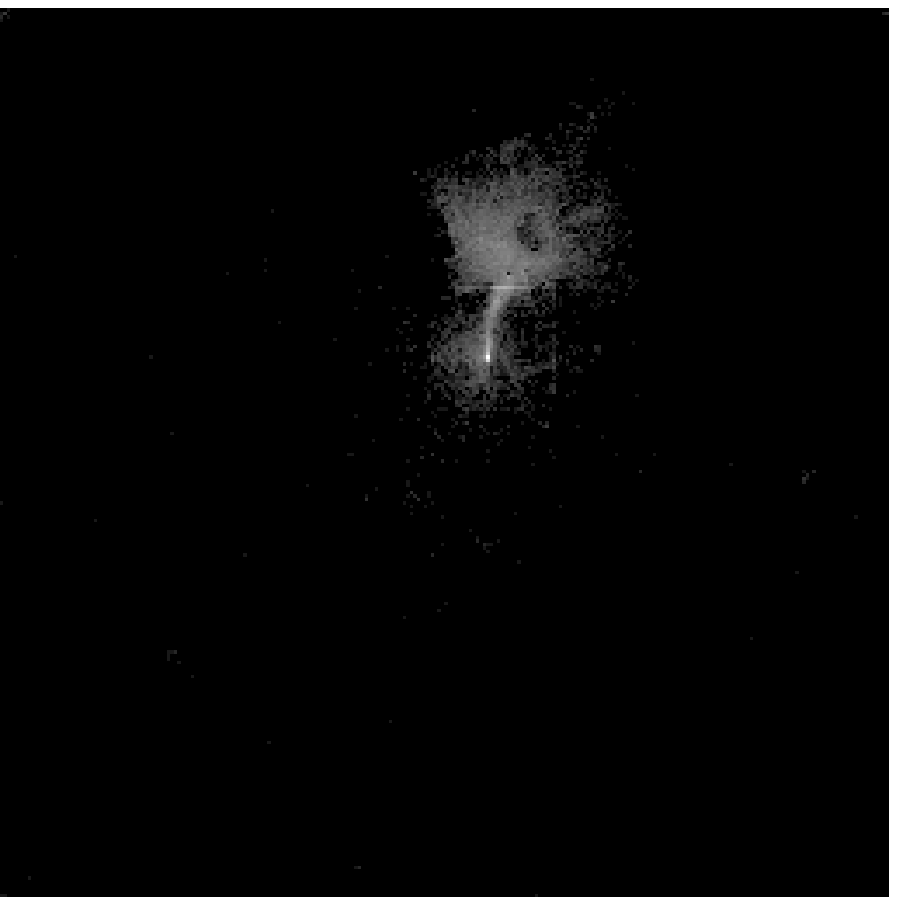}}
\caption{Log-scale choropleth map for number of visits.
Lighter colors means higher value. \label{islands_num_of_visits}}

%\subfigure[Isis: Average Stay Time]{
\subfigure[Isis]{
	\label{isis_avg_pause_time_map}
	\includegraphics[width=.22\textwidth]{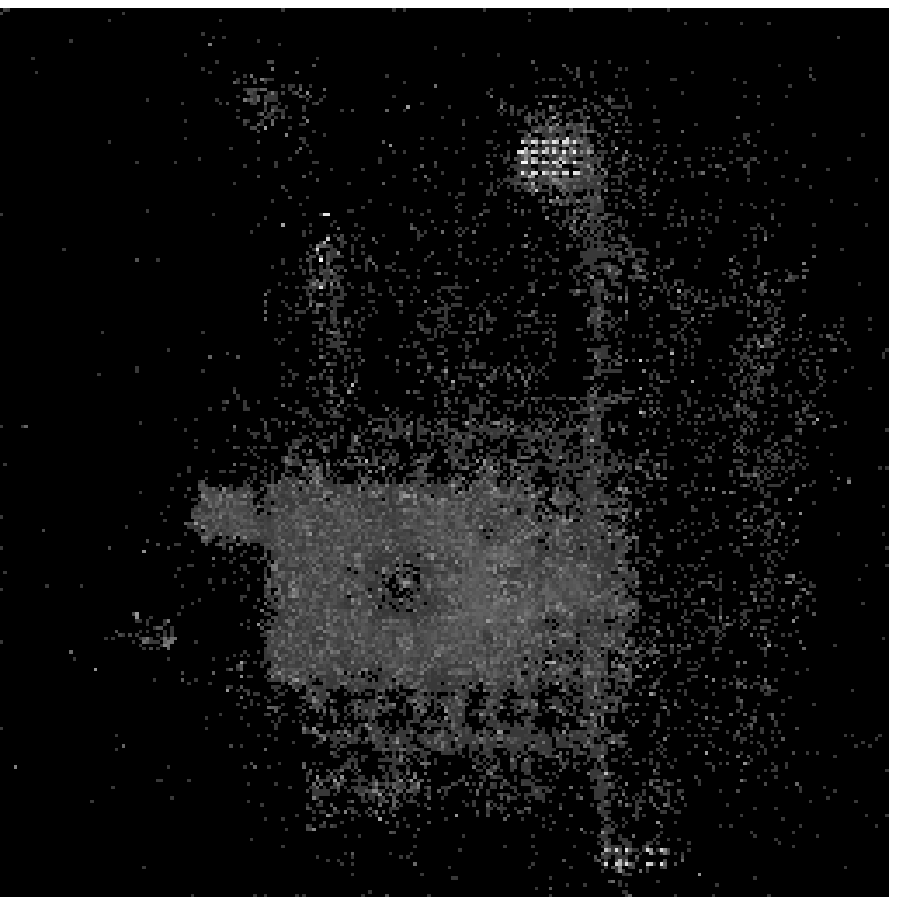}}
%\subfigure[Ross: Average Stay Time]{
\subfigure[Ross]{
	\label{ross_avg_pause_time_map}
	\includegraphics[width=.22\textwidth]{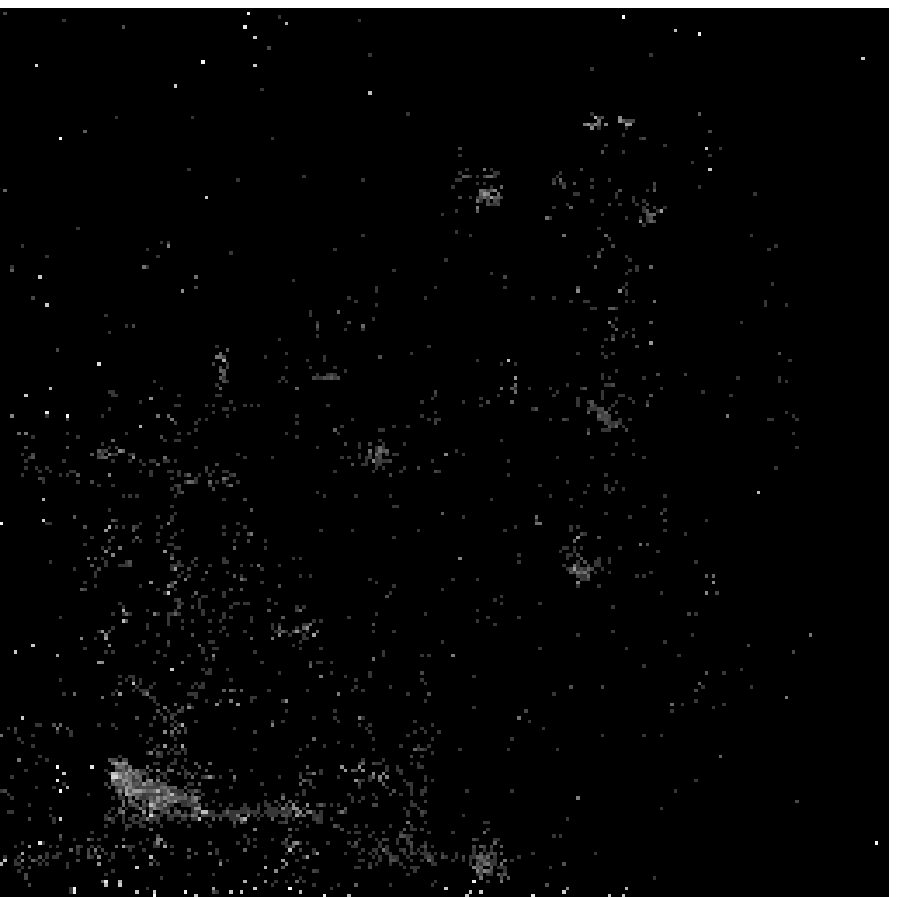}}
%\subfigure[Freebies: Average Stay Time]{
\subfigure[Freebies]{
	\label{freebies_avg_pause_time_map}
	\includegraphics[width=.22\textwidth]{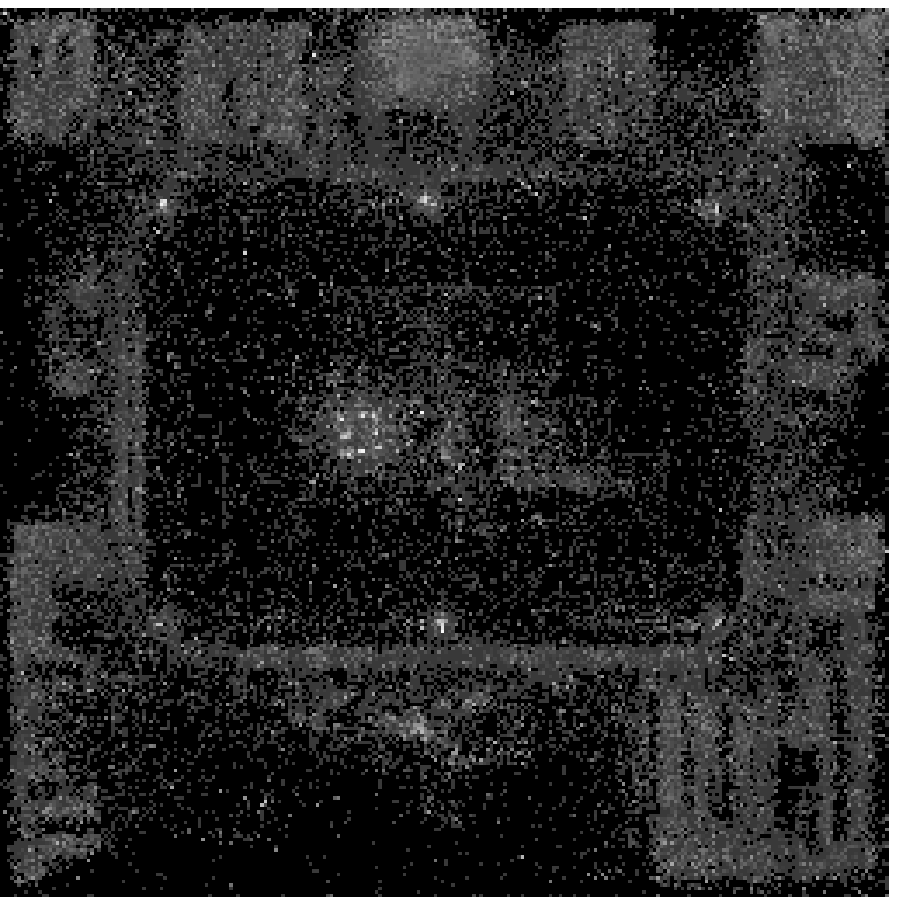}}
%\subfigure[Pharm: Average Stay Time]{
\subfigure[The Pharm]{
	\label{pharm_avg_pause_time_map}
	\includegraphics[width=.22\textwidth]{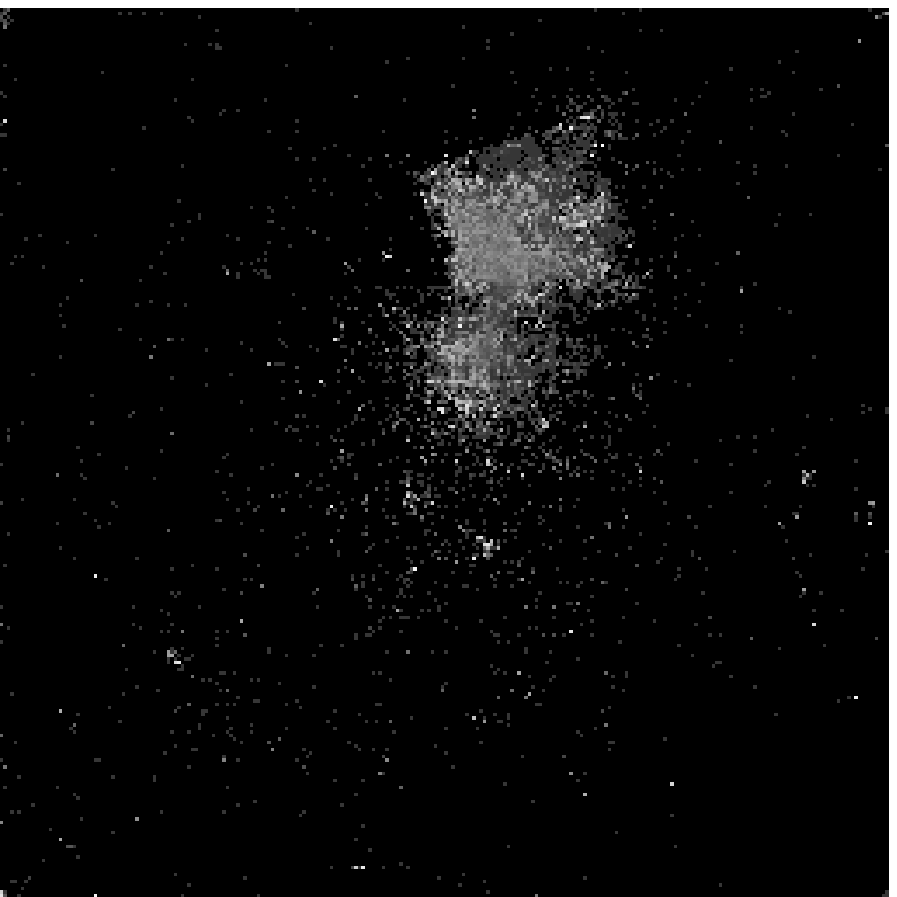}}
\caption{Log-scale choropleth map for average pause time
Lighter colors means higher value.}

\subfigure[Isis]{
	\label{isis_avg_speed_map}
	\includegraphics[width=.22\textwidth]{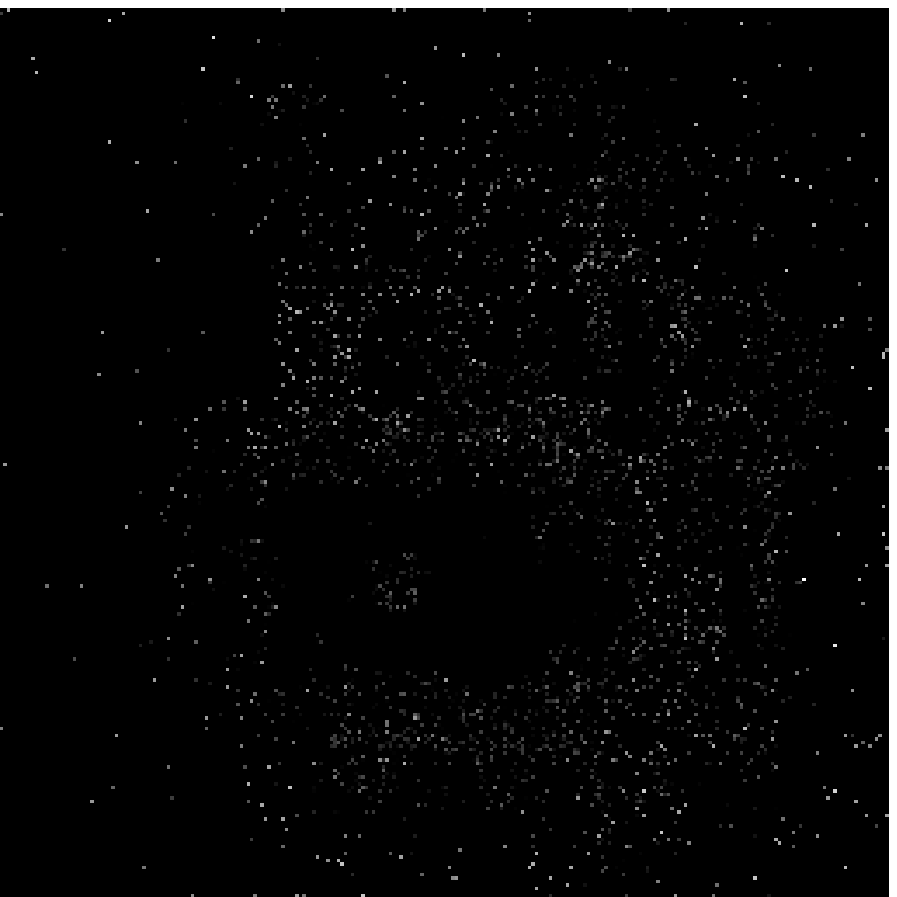}}
\subfigure[Ross]{
	\label{ross_avg_speed_map}
	\includegraphics[width=.22\textwidth]{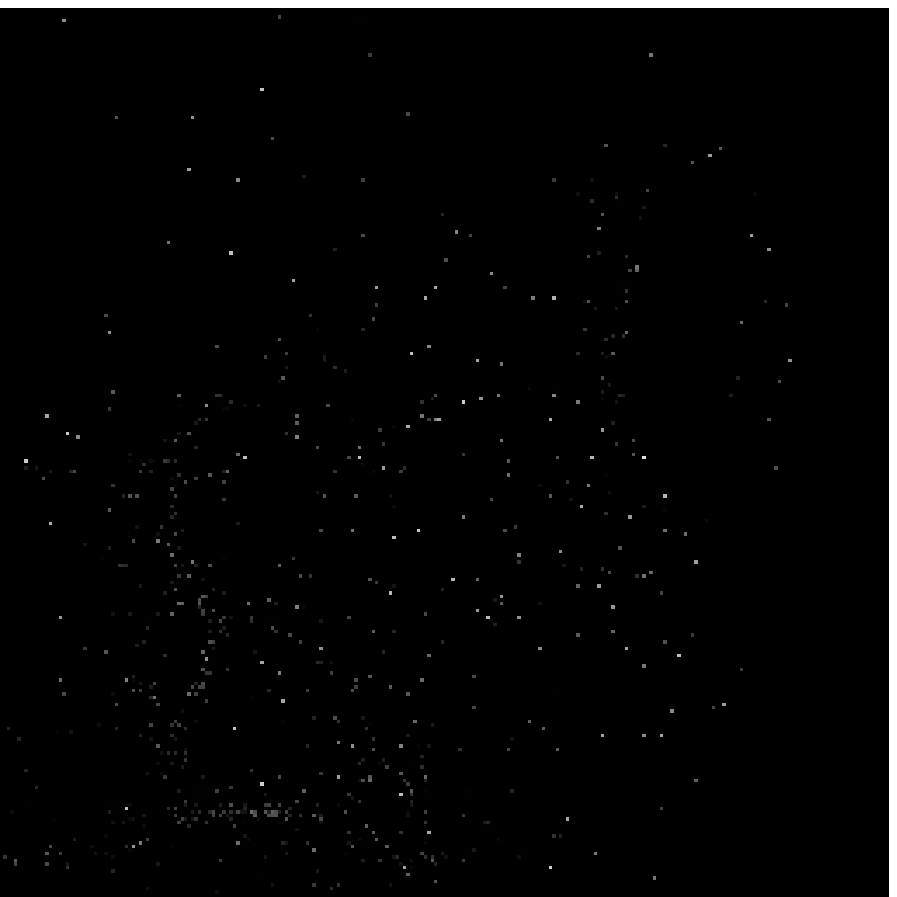}}
\subfigure[Freebies]{
	\label{freebies_avg_speed_map}
	\includegraphics[width=.22\textwidth]{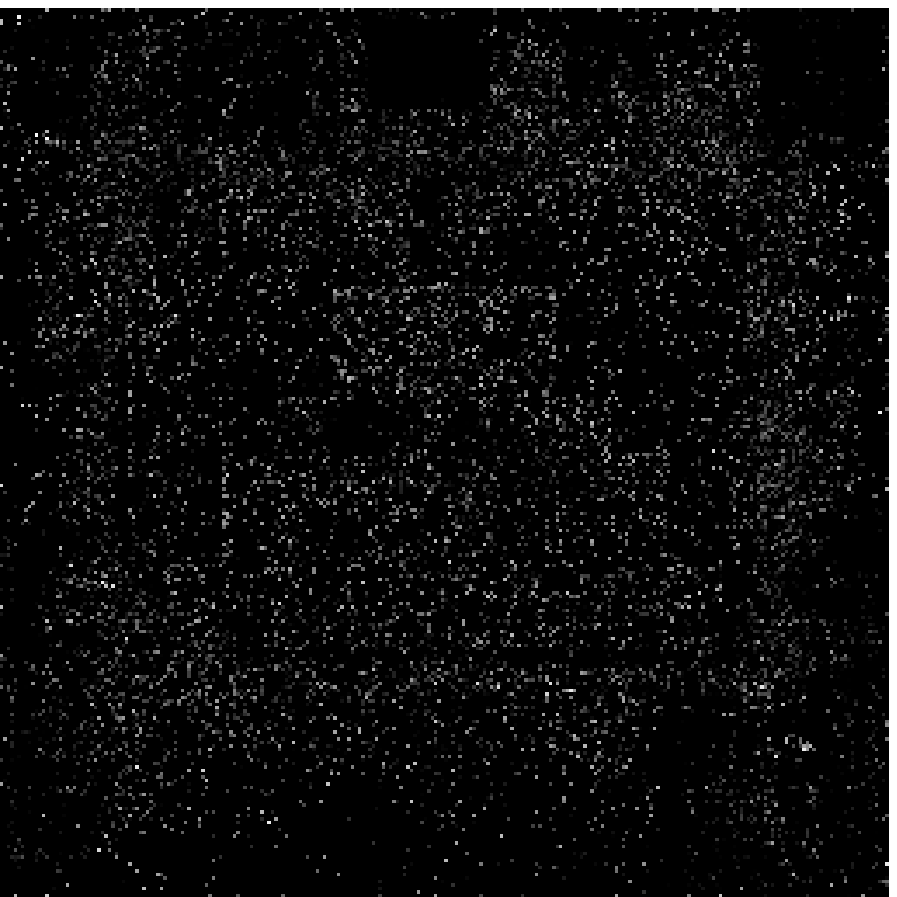}}
\subfigure[The Pharm]{
	\label{pharm_avg_speed_map}
	\includegraphics[width=.22\textwidth]{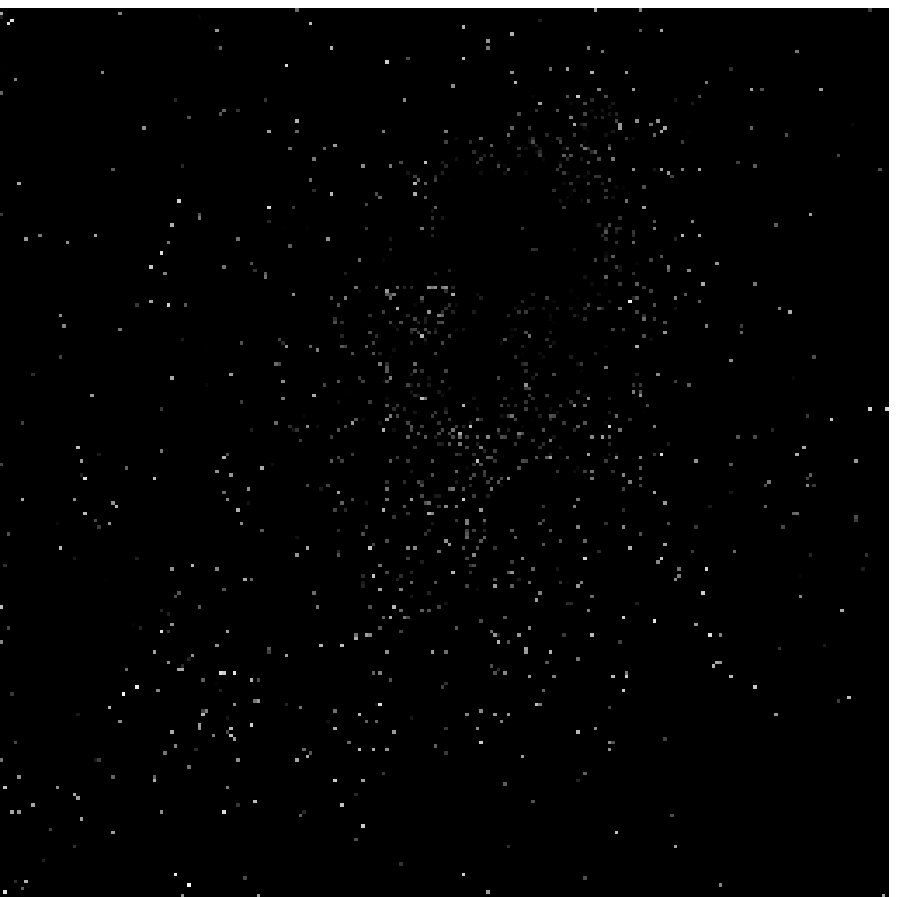}}
\caption{Log-scale choropleth map for average speed.
Lighter colors means higher value.}

\subfigure[Isis]{
	\label{isis_mobility_map}
	\includegraphics[width=.22\textwidth]{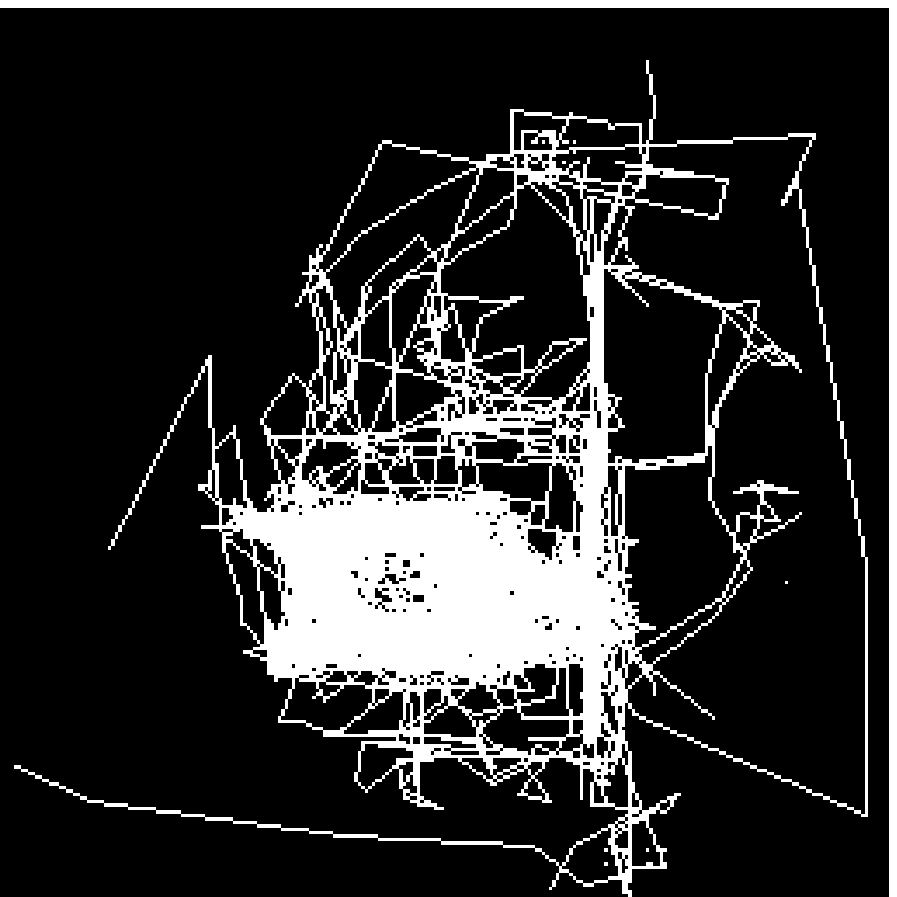}}
\subfigure[Ross]{
	\label{ross_mobility_map}
	\includegraphics[width=.22\textwidth]{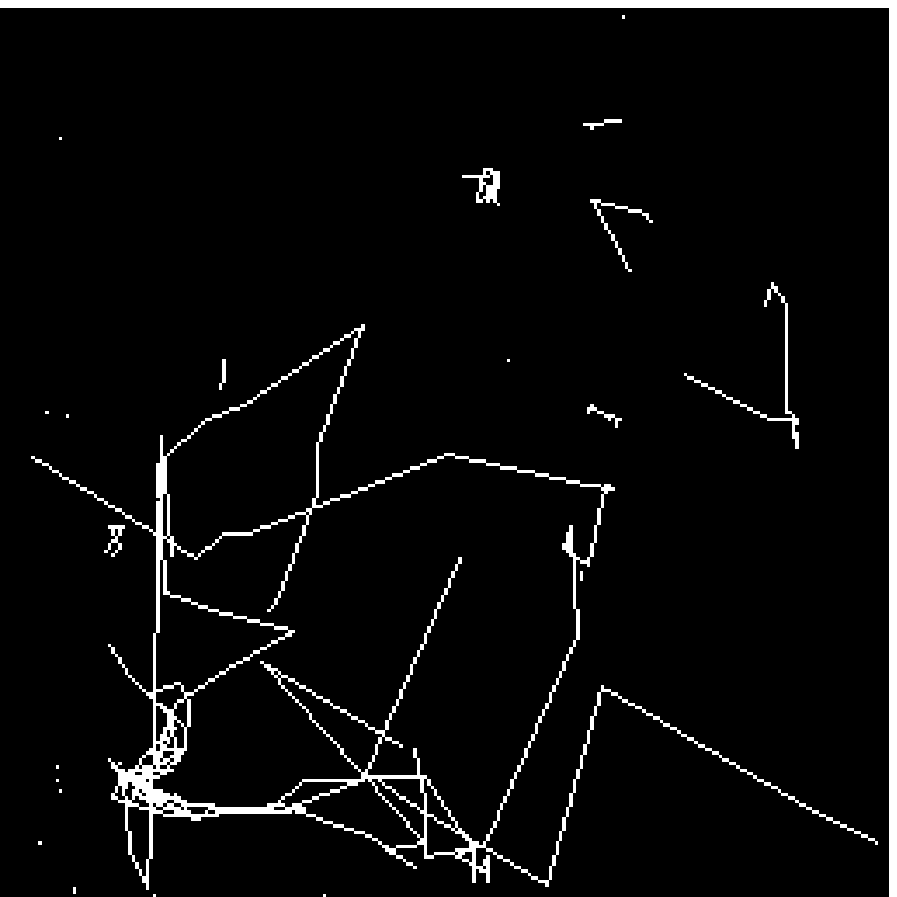}}
\subfigure[Freebies]{
	\label{freebies_mobility_map}
	\includegraphics[width=.22\textwidth]{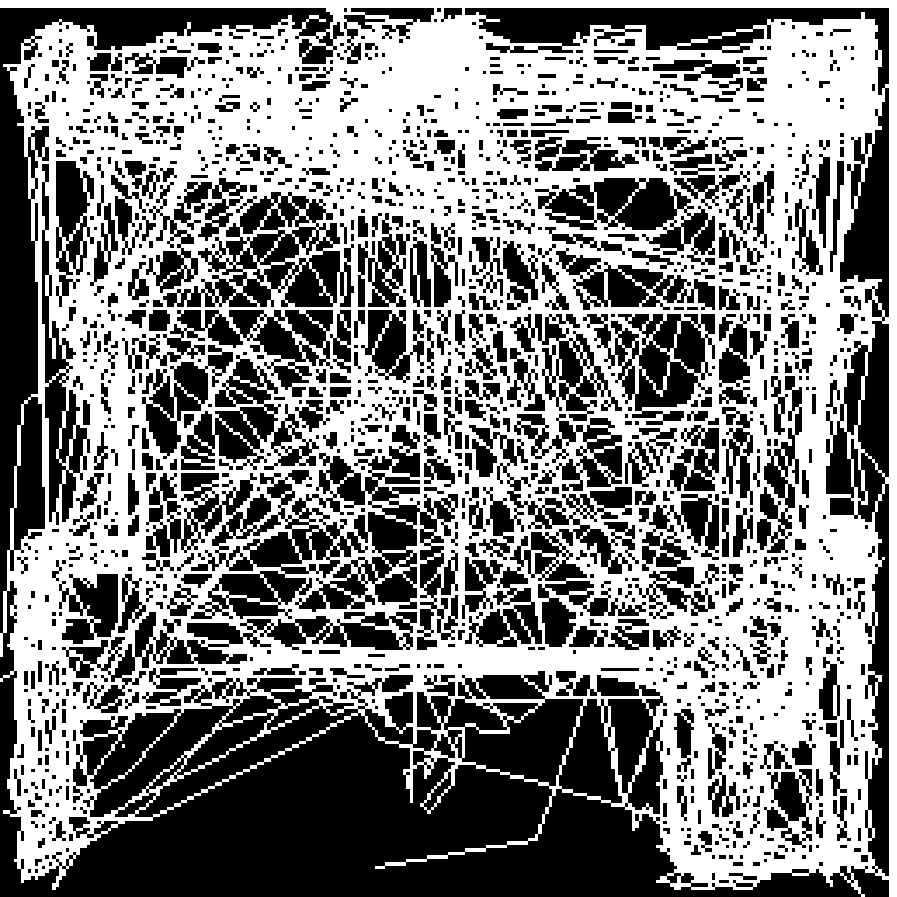}}
\subfigure[The Pharm]{
	\label{pharm_mobility_map}
	\includegraphics[width=.22\textwidth]{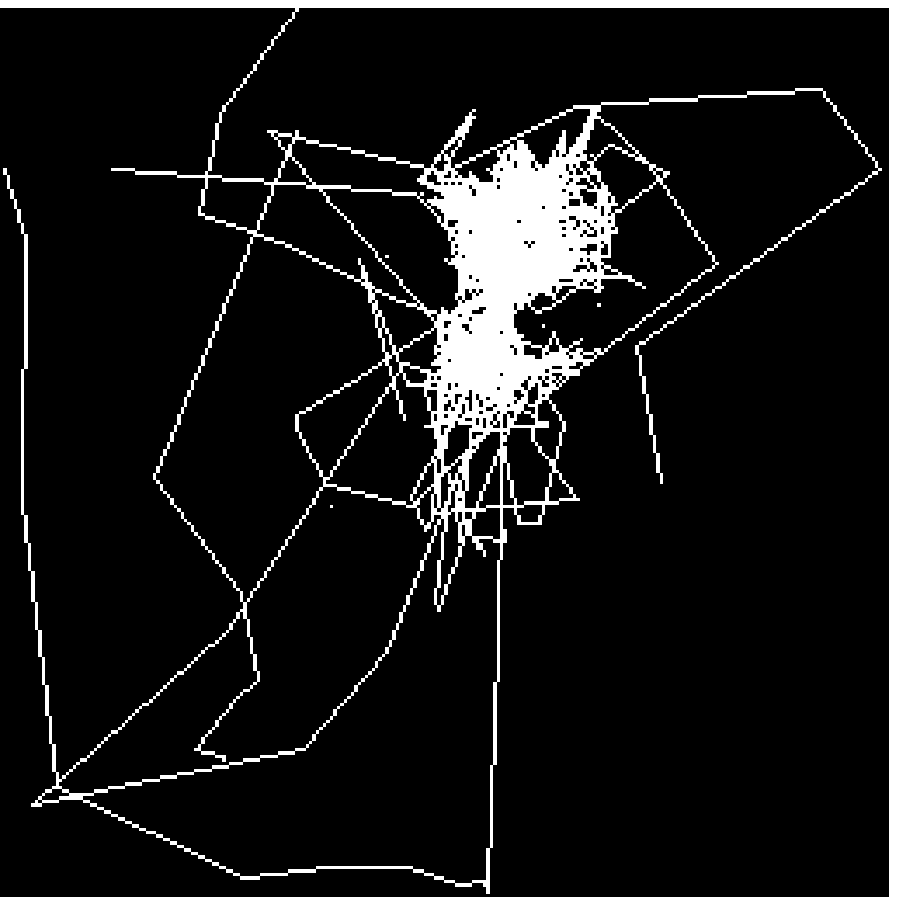}}
\caption{Movement Trails from 6am to 7am \label{islands_mobility_map}}
\end{figure*}

\begin{figure*}[tb]
\centering
\subfigure[Number of Visits]{
	\label{islands_num_of_visits_cdf}
	\includegraphics[width=.31\textwidth]{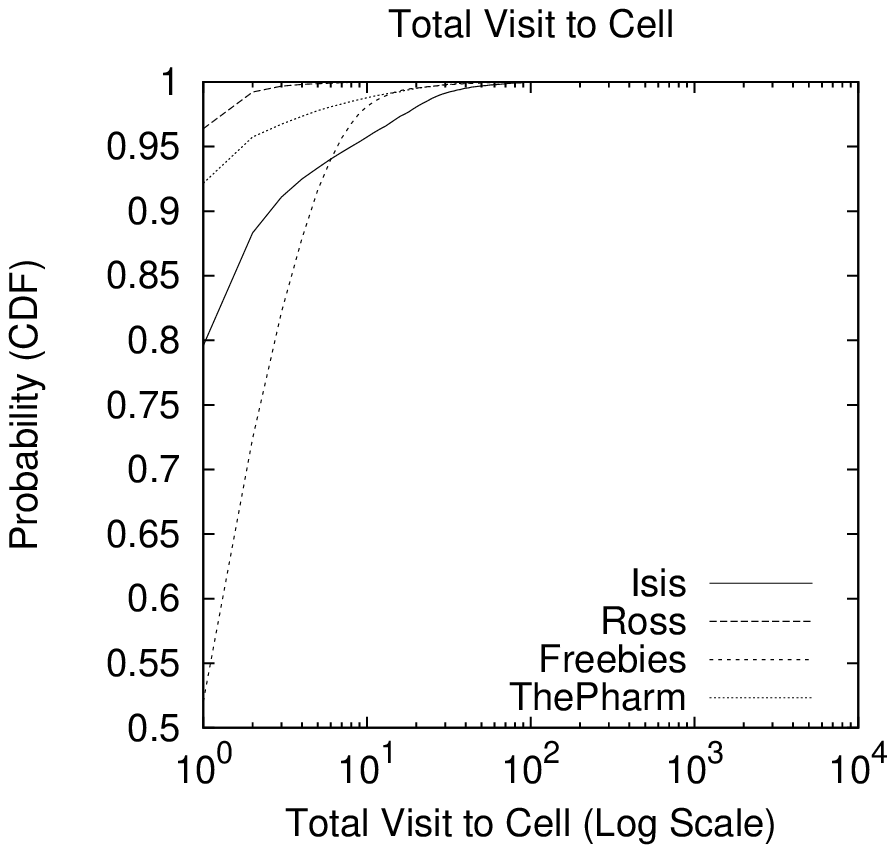}}
\subfigure[Average Pause Time]{
	\label{islands_avg_pause_time_cdf}
	\includegraphics[width=.31\textwidth]{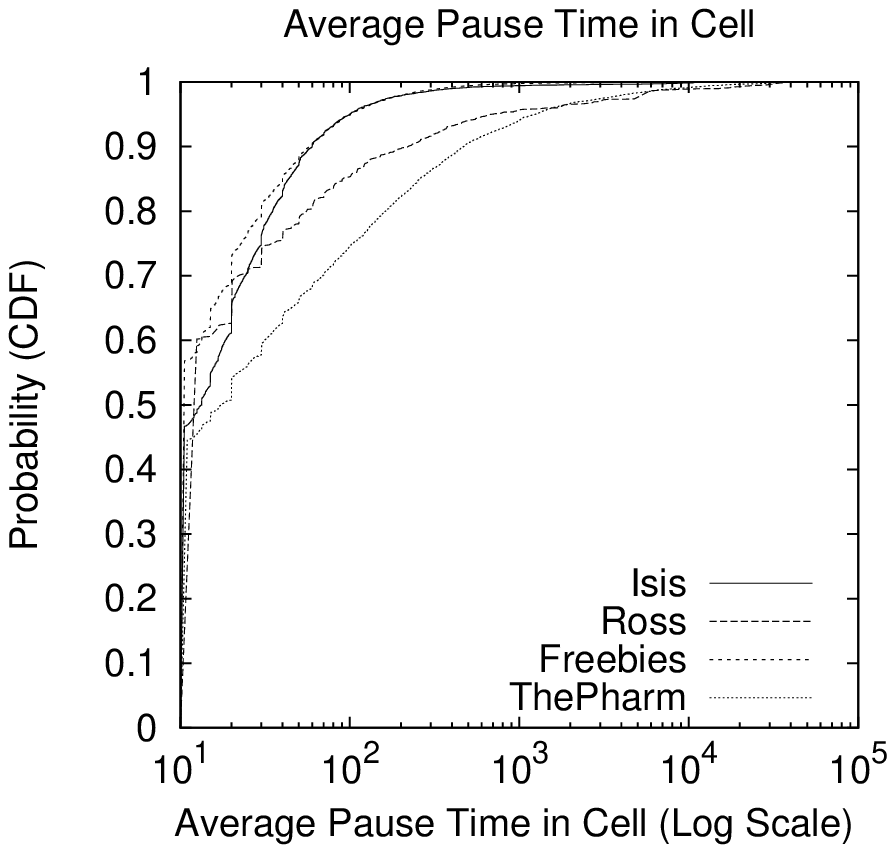}}
\subfigure[Average Speed]{
	\label{islands_avg_speed_cdf}
	\includegraphics[width=.31\textwidth]{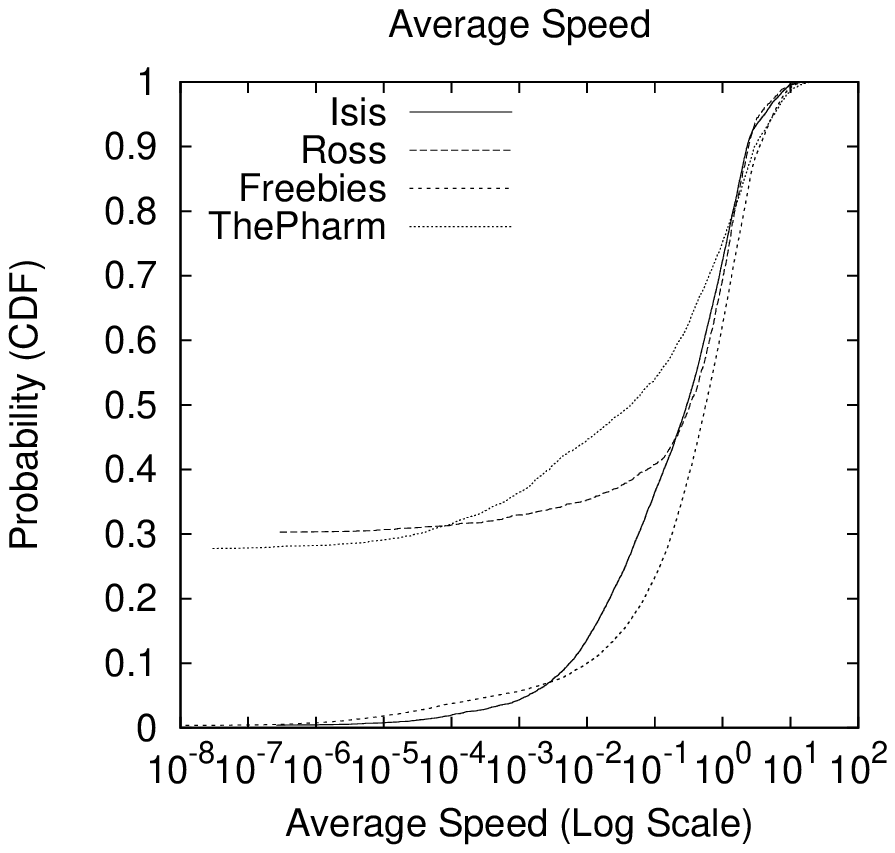}}
\subfigure[Meeting Duration]{
	\label{islands_meeting_contact_time_cdf}
	\includegraphics[width=.31\textwidth]{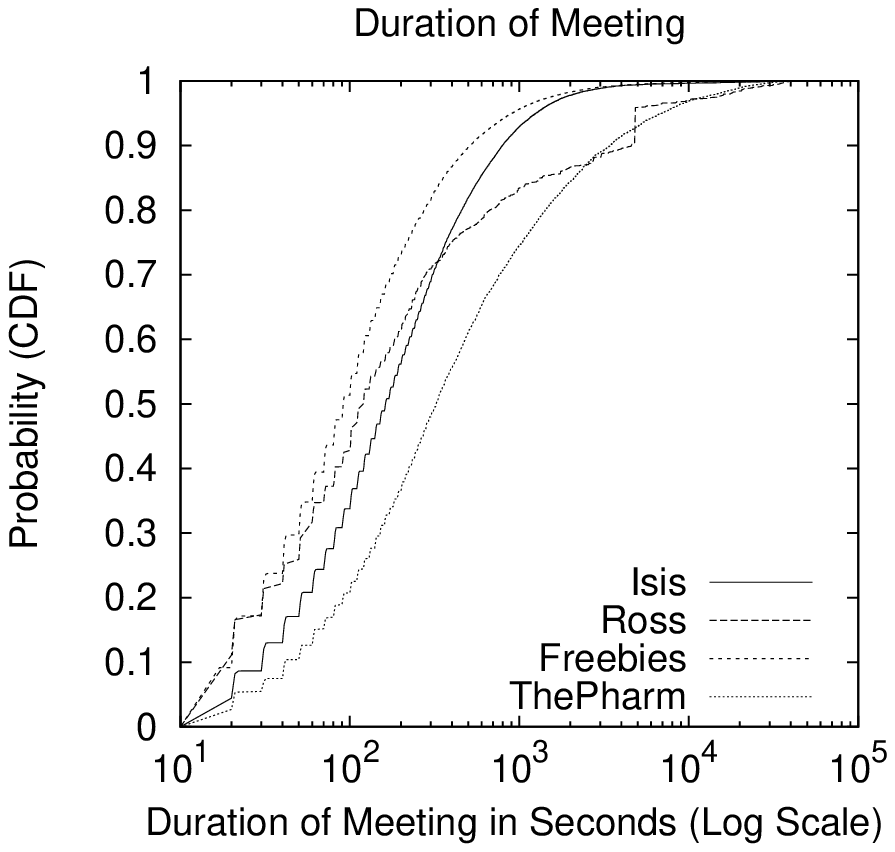}}
\subfigure[Meeting Size]{
	\label{islands_avg_meeting_size_cdf}
	\includegraphics[width=.31\textwidth]{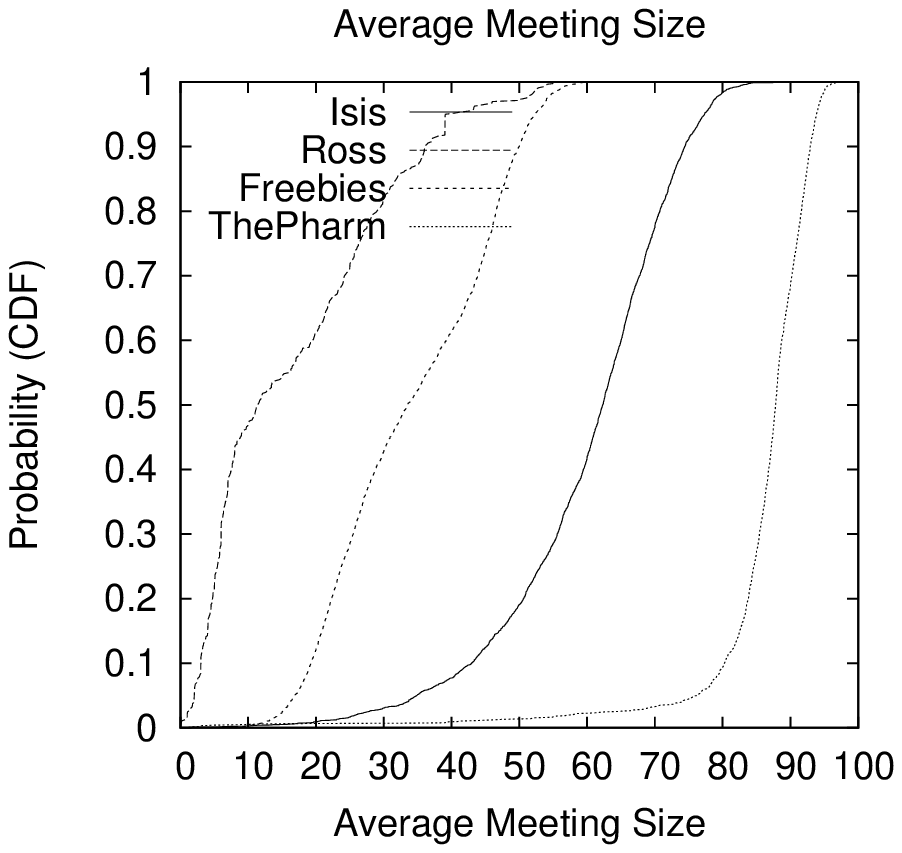}}
\subfigure[Meeting Stability]{
	\label{islands_avg_meeting_churn_cdf}
	\includegraphics[width=.31\textwidth]{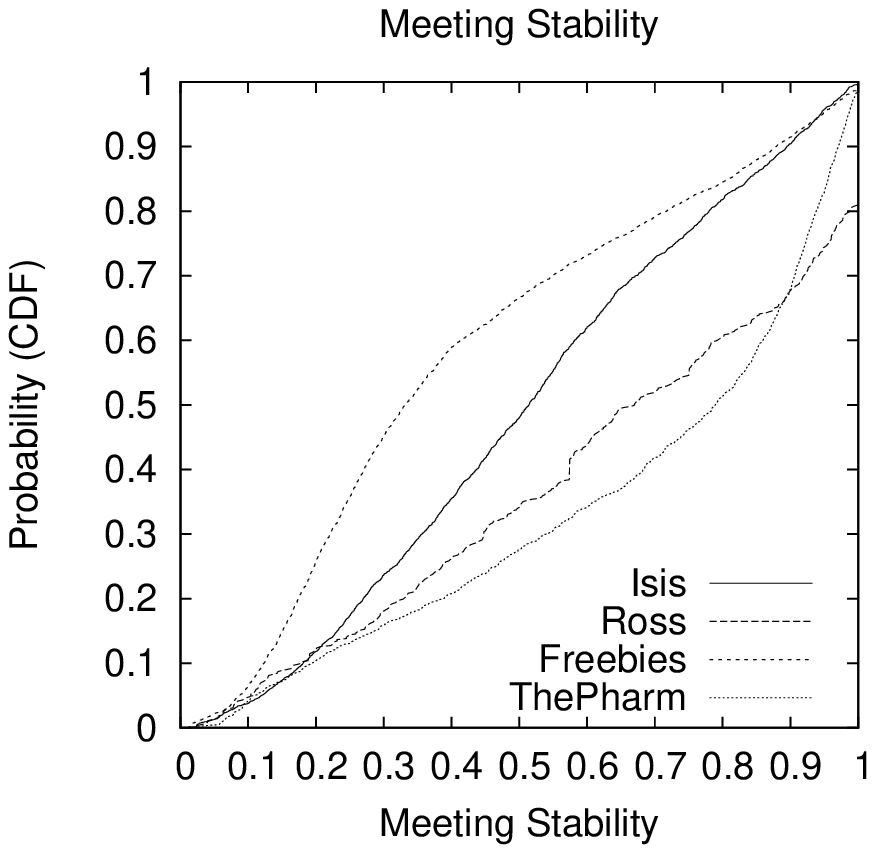}}
\caption{Distributions of Mobility-related Metrics \label{islands_cdf}}
\end{figure*}

We now characterize how avatars move: where they visit, how long do
they pause, how fast do they move, and whether they stay in groups.
We quantize the regions into $256\times256$ equal size 
\textit{cells}, and compute a set of metrics for each cell.  
Figures \ref{islands_num_of_visits}-\ref{islands_mobility_map} %\ref{islands_map} 
show the choropleth maps of the regions for
various metrics.  Figure \ref{islands_cdf}
plots the CDF of the same metrics (x-axis in log-scale).  Figures
\ref{islands_avg_pause_time_cdf} 
and \ref{islands_avg_speed_cdf} shows the distribution over all cells
visited.

\textbf{Number of visits:} An indication of cell
popularity is how many times the avatars visit a cell.  We count
the number of times an avatar enters a cell $c$.
If the same avatar enters and leaves the cell multiple times, it is
counted as multiple visits.  If an avatar logs out and logs in again,
Second Life will place the avatar at the previous position when it 
logs out.  We do not count this entry as a new visit.

Figures \ref{isis_num_of_visits_map} to \ref{pharm_num_of_visits_map} show 
the choropleth map of the number of visits to a cell in log-scale.
The number of
visits to cells is highly skewed.  There are many cells that are not
visited by any avatars in our traces, and a small number of cells 
are visited many times.  Figure \ref{islands_num_of_visits_cdf} 
shows the CDF for number of visits for all four regions.  Freebies
has the most visited cells (52\% are visited at least once), as 
its activities spread to all corners of the region (Figure 
\ref{freebies_num_of_visits_map}).  The regions have 2 (Ross) to 17 (Isis) cells that
are visited more than 100 times, with the most visited cell having
been visited 
2487 times (in Isis).  This cell corresponds to the landing point on
Isis, which can be seen as the bright white spot in Figure
\ref{isis_num_of_visits_map}.  Other regions show similar spots at
their landing points.

\textbf{Average pause time:} Another indication of
the popularity of a cell is the duration an avatar stays in a cell
(until it moves to another cell, teleports to another region,
or logs out).  
The total pause time of a cell $c$ is the sum of the
pause time of all avatars that have ever been to $c$.  Note that since we
log the movement of avatars every 10 seconds, the pause time has
a minimum resolution of 10 seconds.  The average pause time of a
cell is thus total pause time divided by number of visits.
This metric indicates the \textit{stickiness} of a cell.

Figures \ref{isis_avg_pause_time_map} to \ref{pharm_avg_pause_time_map} show 
the choropleth map of the average pause time of a cell in log-scale.
Comparing these maps to 
Figures \ref{isis_num_of_visits_map} to \ref{pharm_num_of_visits_map} shows
there exist cells with high average pause time that are not visited
often (for instance, the group of cells at the top
right quadrant of the map in Figure \ref{isis_avg_pause_time_map}).
The distribution of average pause time is again
highly skewed (Figure \ref{islands_avg_pause_time_cdf}).  On Isis
and Freebies, avatars pause for less than 100 seconds in 95\% of the 
visited cells but 11 and 5 avatars pause more than 3 hours on
Isis and Freebies respectively.  In the camping region, The Pharm, 40
avatars paused more than 3 hours.  The longest average pause time 
observed is just over 14 hours in The Pharm.

\textbf{Average speed in a cell:}  For each pair of
consecutive recorded avatar positions, we note the time between the
records $t$ and the distance traveled $d$.  The average speed of that avatar
is then simply $d/t$.  We consider $d/t$ as the speed of a movement
in cell $c$, where $c$ is the new cell the avatar is in.  The average
speed in a cell is then the average speed over all movements.

The map for average speed is shown in Figures
\ref{isis_avg_speed_map} to \ref{pharm_avg_speed_map},
with the corresponding CDF shown in Figure \ref{islands_avg_speed_cdf}.
Comparing these maps to 
Figures \ref{isis_num_of_visits_map} to \ref{pharm_num_of_visits_map} shows
bright spots (high average speed) outside of the
frequently visited cells.  This observation confirms the intuition
that avatars move quickly (either run or fly in Second
Life) in non-interesting regions but move normally (walk) within
interesting regions.  This is confirmed in Figure
\ref{isis_mobility_map} to \ref{pharm_mobility_map}, which
show the trail of the avatars in a one-hour sub-trace (6am) of the
four one-day traces in Table \ref{tab:regions}.  A long and straight
line means the avatar is moving at a fast speed (including
intra-region teleport).

Figure \ref{islands_avg_speed_cdf} indicates something
interesting: the average speed of about 30\% of the visited
cells have a zero average speed in The Pharm and Ross, implying  
that some avatars never move -- 
likely these users log in and leave the session running,
exploiting camping facilities to earn virtual money
(Note that Second Life leaves the avatar at their previous location
when the user logs in).

\subsection{Contact Patterns}

To characterize the spatial relationship
among avatars, we look at \textit{meetings} among the avatars.  Two
avatars meet if their distance in a region is within a certain 
threshold.  We use 64m as our threshold since this is the default
AoI distance in Second Life.  

\textbf{Meeting Durations:} Figure \ref{islands_meeting_contact_time_cdf}
shows the distribution of meeting duration.  
Meeting durations are long -- over 50\% of the durations are over
82 seconds (for Freebies) and 303 seconds (for The Pharm).  On The Pharm, 9\%
of the meetings are over an hour. 

\textbf{Average Meeting Size:} A closely related measure we compute
is average meeting size over time.  For every 10 seconds, we computed
the \textit{meeting size} -- the number of avatars within their AoI.  
We then compute, for each avatar, the average meeting size over time.  
We found that the meeting size is generally large in the regions we
studied.  On The Pharm, the
average meeting size for avatars is above 40 for 99\% of the avatars.
Even on Ross, a medium popularity region, the median meeting size is 11.4.

\textbf{Meeting Stability:} To see how the avatars in a meeting change
over time, for each avatar $a$, we take the ratio of average meeting size
over the number of unique avatars ever met by $a$ in the region.  We
call this ratio \textit{meeting stability}.  If an avatar's stability is 1, then 
the avatar always meets with the same set of avatars while she is
in the region.  The distribution of this ratio is shown in Figure
\ref{islands_avg_meeting_churn_cdf}.  
Avatars in 
The Pharm and Ross have a high meeting stability, with 50\% or more
having a stability of 0.67 and 0.79 respectively.  19\% of avatars have
a stability of 1.0 in Ross.  Avatars in Freebies show much more
dynamic behavior, with a median stability of only 0.33.

\subsection{Temporal Variation}
We are interested in seeing if the general observations we made above
change over time.  We analyze a four-day trace from Isis, hour-by-hour.
We pick the number of visits to a cell as the metric to study, as it
is a good indication of whether a cell is popular.  For each cell, we
sample the number of visits to that cell in each hour, and computed
the standard deviation of the samples.  

We found very little variation in the number of visits to the cells,
hour-by-hour, over the course of four days.
99\% of the cells have a standard deviation less than 1.05.
The largest standard deviation, 20, is observed at the landing
point.

\section{Implications}
\label{sec:implications}

In this section, we discuss our observations from the traces and
how they relate to existing research in NVEs.  

\subsection{Peer-to-peer NVEs}

Centralized server architectures, such as those employed by Second
Life, do not scale well to a large number of players.
This challenge of scalability
has motivated research into alternative architectures,
one of which is the peer-to-peer architecture, where
clients communicate directly with other clients through an overlay,
without going through the server \cite{Gautier98,Hu06}.  A client may 
also share some responsibility of the server (such as maintaining 
states) \cite{Bharambe06,Lu04}.  

Our traces show evidence of high churn rate, averaging about one
every two minutes on the popular regions, and could have drastic 
effect on the efficacy of peer-to-peer NVEs.  Such high churn rates
imply that the system has to continuously configure the peer-to-peer 
overlay and sufficient redundancy 
needs to be built-in to prevent loss of information \cite{Bharambe06}.
Our analysis shows that the stay time is highly skewed.
This observation supports peer-to-peer NVEs that employ super-nodes to
store states and manage other peers \cite{Lu04,Chen06}.  The avatars who 
stayed for a long time in a region may be good candidates to be super nodes.

Our traces also suggest a novel and interesting way to identify the
potential super nodes.  We observe that there are cells within the regions that
are ``sticky'' -- avatars tend to pause within these cell for much longer 
time than other cells.  Thus, avatars that pause at these sticky
cells are more likely to pause for a long time.

Many P2P NVE schemes build an overlay by connecting peers
within an AoI as neighbors
\cite{Keller03,Hu06,Chen05MIP}.  Our traces support this design in
regions with low popularity and low mobility (such as Ross and The
Pharm), where the meeting stability is relatively high.  For regions
such as Isis and Freebies, where the AoI neighbors change frequently,
reducing overhead in establishing and tearing down connections among 
neighbors remains a challenge.

\subsection{Zone Partitioning and Load Balancing}
\label{sec:zone}

Besides P2P architectures, another architecture that has been proposed
to improve the scalability of NVEs is to employ a cluster of servers.  
The game world is divided into zones, each 
managed by a server.  This architecture is similar to what Second
Life employs today.  The research challenge is to make the game 
world seamless by making zoning transparent to the users.  

A consequence of seamless zone partitioning is that we can 
dynamically repartition the zones to balance the load on the servers
\cite{Vleeschauwer05,Chen05,Lui02,Chertov06}.  Two important factors
that affect any load balancing schemes are population in a zone, 
and movement across zones.

Our traces show that, unsurprisingly, the spatial distribution of
avatars is far from uniform, especially for Isis and The Pharm.   
This observation supports the need for more sophisticated partitioning
scheme, beyond grid-based partitioning.

We also observed from our traces that avatars tend to move quickly
in lowly populated, non-interesting areas within a region.  This 
creates another issue in which such fast-moving avatars have to be
handed off from one server to another as they move across the zones,
increasing server overheads.  The existence of such behavior also points
to the importance of load balancing schemes that dampen sensitivity 
to minor increase in load -- to avoid frequent triggers of the
load balancing mechanism due to such fast-moving avatars.  
Our traces
show that the popularity of cells does not show much variation over 
time, suggesting that dynamic load re-balancing of zone would occur only
rarely.

\subsection{Interest Management}

Interest management techniques suppress updates from
one avatar to another avatar if the two avatars
are deemed to be irrelevant to each other.  For peer-to-peer
architectures, there is no centralized authority that stores the
states of all avatars.  Researchers have previously proposed novel,
distributed schemes to determine the interest between two avatars
\cite{Makbily99,Steed05}.  The ideas behind these schemes are that,
two avatars exchange their locations and each computes a safe zone\footnote{the term
zone here has no relationship to that in Section \ref{sec:zone}},
based on occlusions and visibility information, where
they can remain in without needing to update each other.  If one
avatar moves out of this safe zone, they exchange their location again
and recompute the safe zones.

Our traces mean both good news and bad news
for these schemes.  On one hand, we found that in interesting cells,
where there are typically many occlusions, avatars move slowly. 
Thus avatars need to update each other rarely.  On the other hand, 
in non-interesting cells, there are few occlusions and 
avatars move quickly, and thus need to update their locations to each
other frequently.

Since determining visibility between avatars can be expensive, another
popular technique for determining relevancy between two avatars is
to use the AoI -- two peers update each other as long as
their avatars are in the other's AoI.  Our traces indicates 
large average meeting size
(Figure
\ref{islands_avg_meeting_size_cdf}),
suggesting high overhead in exchanging updates for this technique.

\subsection{Mobility Modeling}

Our traces are useful in designing and verifying new mobility
models for NVEs.  We observed that on popular regions such as
Isis and Freebies, avatars tend on congregate in interesting places
and move at a slower speed.  They also move faster in non-interesting
places, perhaps exploring and looking for
interesting things, or moving from one interesting place to another
(See Figures \ref{isis_mobility_map}-\ref{pharm_mobility_map}).
This observation suggests that simple mobility models such 
as random walk and random way-point \cite{Johnson96} are insufficient 
in modeling mobility of all regions in NVEs such as Second Life.

The movement within high density areas in a region seems to suggest
a pathway model -- where avatars move along constrained paths
(such as corridors or bridges) and visit various rooms \cite{Tian02}.  
This model, however, does not account for the high speed movements when
avatars move in non-interesting cells.  Thus, our measurements 
seem to suggest that a hybrid mobility model that incorporates both
random way-point mobility model (for outdoor) and pathway mobility 
model (for indoor) would be more suitable. 
Our analysis also suggests the mobility model should incorporate
skewed distributions in movement speed and pause time.

\subsection{Prefetching}

Prefetching is commonly used to reduce the 
object access latency in NVEs \cite{Chan01,Chim98,Lau01,Park01}.  
The key to successful prefetching is predicting accurately
which objects are needed, so that bandwidth is not wasted
in retrieving objects that will not be eventually used. 

Second Life prefetches the data within a circular region of an
avatar.  We found from our Isis and Freebies traces, however,
that avatars only spent 18\% of their time rotating around a
point.  

Our extended analysis of the multiple day traces
reveals that the popularity 
of the cells in a region do not change over days, or even 
over hours.  This observation suggest that we can use short-term,
historical information about the popularity of a cell as
an input to the prefetching algorithm to help with predictions.  

\subsection{User-based Caching}
Our traces show that even within a day, there are multiple
revisits to the same region by the same avatar.  This pattern
supports user-based caching in Second Life.  It also suggests
the possible benefits of \textit{region-aware caching}, which
considers access patterns to 
a region, in addition to access patterns to objects, in the
cache replacement algorithm -- so that if a user
is hopping from region to region, objects from a region they 
repeatedly visit are not evicted from the cache when they
visit other region.

\section{Conclusion}
\label{sec:conclude}
This paper presents our effort in collecting and analyzing 
large traces of avatar mobility in Second Life.  We focus 
on four regions with different characteristics in this paper.  
It would be interesting to capture such traces on regions 
which hold transient events (such as parties), where popularity 
could vary both spatially and temporally.

In this paper, we computed baseline metrics related to system 
design in NVEs.  One could mine the traces for answers to other 
interesting questions (such as whether an avatar tends
to move towards cells with other avatars).
We plan to continue analyzing our traces to reveal interesting 
patterns in avatar mobility.  

We also discussed how our findings could
impact many other aspects of NVE design and suggested several
potentially beneficial ideas
(e.g., identifying super-nodes, region-aware caching, and 
popularity-based prefetching).  We plan to investigate these
ideas in depth and evaluate their effectiveness using our traces.

\section*{Acknowledgment}
We thank Vikram Srinivasan of Bell Labs Research India for 
useful discussions throughout this project.

\bibliographystyle{plain}
\bibliography{secondlife-mobility}
\end{document}